\documentclass[%
 aip,
 amsmath,amssymb,
preprint,%
]{revtex4-1}

\usepackage{graphicx}
\usepackage{dcolumn}
\usepackage{bm}
\usepackage[inline]{enumitem}
\usepackage[caption=false]{subfig}
\usepackage{hyperref}
\usepackage{xcolor}
\usepackage{multirow}

\usepackage[utf8]{inputenc}
\usepackage[T1]{fontenc}
\usepackage{mathptmx}
\usepackage{booktabs}

\newcommand{\add}[1]{\textcolor{black}{{}#1}}

\begin{document}

\preprint{AIP/123-QED}

\title{A General Automatic Method for Optimal Construction of Matrix Product Operators Using Bipartite Graph Theory}

\author{Jiajun Ren}
 \email{renjj@mail.tsinghua.edu.cn}
 \affiliation{MOE Key Laboratory of Organic OptoElectronics and Molecular
 Engineering, Department of Chemistry, Tsinghua University, Beijing 100084, People's Republic of China }
\author{Weitang Li}
 \affiliation{MOE Key Laboratory of Organic OptoElectronics and Molecular
 Engineering, Department of Chemistry, Tsinghua University, Beijing 100084, People's Republic of China }
 \author{Tong Jiang}
 \affiliation{MOE Key Laboratory of Organic OptoElectronics and Molecular
 Engineering, Department of Chemistry, Tsinghua University, Beijing 100084, People's Republic of China }
\author{Zhigang Shuai}%
\affiliation{MOE Key Laboratory of Organic OptoElectronics and Molecular
 Engineering, Department of Chemistry, Tsinghua University, Beijing 100084, People's Republic of China 
}%

\date{\today}

\begin{abstract}
Constructing matrix product operators (MPO) is at the core of the modern density matrix renormalization group (DMRG) and its time dependent formulation. For DMRG to be conveniently used in different problems described by different Hamiltonians, in this work we propose a new generic algorithm to construct the MPO of an arbitrary operator with a sum-of-products form based on the bipartite graph theory. We show that the method has the following advantages:
\begin{enumerate*}[label=(\roman*)]
\item It is automatic in that only the definition of the operator is required;
\item It is symbolic thus free of any numerical error;
\item The complementary operator technique can be fully employed so that the resulting MPO is globally optimal for any given order of degrees of freedom;
\item The symmetry of the system could be fully employed to reduce the dimension of MPO.
\end{enumerate*}
To demonstrate the effectiveness of the new algorithm, the MPOs of Hamiltonians ranging from the prototypical spin-boson model and Holstein model to the more complicated ab initio electronic Hamiltonian and the anharmonic vibrational Hamiltonian with sextic force field are constructed. It is found that for the former three cases, our automatic algorithm can reproduce exactly the same MPOs as the optimally hand-crafted ones already known in the literature.
\end{abstract}

\maketitle

\section{Introduction}

The density matrix renormalization group (DMRG) method originally proposed by White to solve the electronic structure of one-dimensional strongly correlated lattice models~\cite{white1992density} has made great progress in quantum chemistry in the last decade and has been widely recognized as a state-of-the-art method for problems with a large active space.~\cite{chan2011density,ren2016inner,ma2017second,yanai2017multistate,li2017spin,luo2018externally,baiardi2020density,brandejs2020toward} In addition to the electronic correlation, DMRG also shows great potential to solve the vibrational correlated problems.~\cite{rakhuba2016calculating,baiardi2017vibrational,baiardi2019optimization} More recently, the time dependent (TD) formulation of DMRG called TD-DMRG attracts a lot of attention and quickly emerges as an efficient and ``nearly exact'' method for quantum dynamics in complex systems. TD-DMRG has been used to simulate the spectroscopy and quantum dynamics, including not only  electron dynamics~\cite{ronca2017time,frahm2019ultrafast} but also electron-vibrational correlated dynamics.~\cite{greene2017tensor,yao2018full,ren2018time,kurashige2018matrix,baiardi2019large,xie2019time,li2020numerical,Li2020finite} 
\add{For high-dimensional quantum dynamics, the multi-configuration time-dependent Hartree (MCTDH) method has long been considered as the gold standard.~\cite{meyer1990multi,beck2000multiconfiguration} However, it is limited by the exponential growth of computational cost with the system size -- the curse of dimensionality. The multilayer MCTDH (ML-MCTDH) overcomes this limitation and has been successfully applied to simulate dynamics of model systems with thousands of degrees of freedom (DoF).~\cite{wang2003multilayer} Like ML-MCTDH, (TD-)DMRG could also achieve arbitrarily high accuracy with only polynomial computational effort. It has been demonstrated in a number of models with hundreds of DoFs to have the same accuracy as ML-MCTDH.~\cite{ren2018time,xie2019time,yao2018full,kurashige2018matrix,larsson2019computing}}

The recent rapid advances in quantum chemistry DMRG can be attributed to the formulation of DMRG as matrix product state (MPS)~\cite{ostlund1995thermodynamic} and the corresponding operator could be represented as matrix product operator (MPO).~\cite{crosswhite2008finite}
The introduction of MPS and MPO not only establishes a rigorous mathematical foundation of DMRG, but also makes the algorithm more powerful and convenient.~\cite{schollwock2011density} Furthermore, it also opens the door to the development of more general tensor network states (TNS) such as tree tensor network states (TTNS)~\cite{nakatani2013efficient,gunst2018t3ns} and projected entangled pair states (PEPS).~\cite{li2019generalization} 
\add{DMRG is actually a special type of TNS with an one-dimensional matrix product ansatz, which is mathematically known as a tensor train (TT) format.
Interestingly, from the TNS point of view, ML-MCTDH has a TTNS wavefunction ansatz with all physical DoF (primitive basis) at the leaf-node (lowest layer), which is mathematically called hierarchical Tucker format. In this sense, the ansatz of DMRG and ML-MCTDH are both low-rank approximations to the exact high-rank wavefunction, although historically they are independently developed in different research fields~\cite{white1992density,meyer1990multi,wang2003multilayer}. }
The modern formulation of the DMRG algorithm based on MPS and MPO is usually called the second generation DMRG algorithm,~\cite{keller2015efficient} which could be seamlessly combined with the variational principle to obtain the ground state and the time dependent variational principle to carry out the time propagation.~\cite{haegeman2016unifying,li2020numerical} In addition, the exact global arithmetic, such as additions $\Psi_1 + \Psi_2, \hat{O}_1 + \hat{O}_2$ and multiplications $\hat{O} \Psi, \hat{O}_1 \hat{O}_2$, are only possible based on MPS and MPO. In this new formulation, the starting point is to construct the MPO representation of the Hamiltonian and all the other required operators as an input to the rest DMRG calculations. 

\add{Generally speaking, there are two different types of operators. One is the analytical operators, such as the ab initio electronic Hamiltonian in quantum chemistry, the nuclear kinetic energy operator in an appropriate set of coordinates and most of the physical and chemical model Hamiltonians. In addition, these analytical operators are commonly in a sum-of-products (SOP) form. The other type is the potential energy operator of real molecules met in molecular nuclear Schr\"odinger equation. More specifically, the potential energy ``operator'' here is a complex $N$-dimensional potential energy surface (PES) $V(\bm{q}) = V(q_1, q_2, \cdots q_N)$ and has no analytical forms. The potential energy (or energy derivative) at a specific structure $\bm{q}$ could be calculated by electronic structure calculation. 
For small-sized molecules with several atoms, very high accurate PESs are usually constructed globally by fitting and interpolating the available dataset of ab initio data points. The recent developed algorithm based on the neural network (NN) has made great progress in the direction.~\cite{raff2012neural,jiang2013permutation} For medium- and large-sized molecules, constructing full-dimensional global PESs is not even possible. The local PES around the equilibrium or saddle point could be expanded as a Taylor series with the high order energy derivatives. Though the Taylor expansion of PES has several known limitations such as that it could not describe double well potential and large amplitude motion, it is still very useful to calculate the anharmonic frequency of semi-rigid molecules and obtain a more accurate IR/Raman spectrum beyond the harmonic approximation.~\cite{barone2014fully,barone2005anharmonic,barone2012toward}
For most numerical methods to solve the nuclear Schr\"odinger equation such as (ML-)MCTDH, one difficulty is to calculate the matrix element such as $\langle\Psi| V(\bm{q})|\Psi \rangle $, which is an $N$-dimensional quadrature problem. To reduce the cost, it is preferred to decompose the potential into an SOP form. In this way, the matrix elements could be calculated as a sum of product of $N$ one-dimensional quadrature. The Taylor series expansion of the local potential apparently has an SOP form. For a general PES, 
Potfit~\cite{jackle1996product} and the more efficient multigrid Potfit method~\cite{pelaez2013multigrid} could decompose the PES numerically into a Tucker format from the energy grid points, which is suitable to the MCTDH calculation. The recently developed multilayer Potfit could  integrate more effectively with ML-MCTDH.~\cite{otto2014multi} 
Besides the Potfit-like methods, in the NN algorithm to fit the PES, if the activation function is an exponential function instead of the common hyperbolic tangent or sigmoid function, the NN with a single hidden layer also gives an analytical SOP form.~\cite{manzhos2006using} 
In addition to the SOP form, one of the other widely used methods to overcome the $N$-dimensional quadrature is called $n$-mode representation ($n$-MR), in which the PES is expanded as a sum of one-mode potential, two-mode potential and so on, expecting that the series could converge with a small number of terms.~\cite{carter1997vibrational} Thus, only low dimensional quadrature is needed. If necessary, each term in $n$-MR could be further fitted as a sum of products of analytical functions such as polynomial and Morse types for each individual mode.~\cite{klinting2018employing}
It is also worth mentioning that several methods could directly use the $N$-dimensional PES, like the MCTDH combined with correlation discrete variable (CDVR)
representation~\cite{manthe1996time} and its multilayer generalization~\cite{manthe2008multilayer} proposed by Manthe and the collocation method proposed by Carrington \textit{et al}.~\cite{avila2013solving} }

\add{In this work, we focus on the construction of MPO for those operators that have an SOP form by definition or have been transformed into an SOP form by fitting a high-dimensional function (discrete points) as introduced above.}
For the same operator, the form of MPO could be completely different as long as the final product is correct. However, a more compact MPO will save computational cost in practice.
In order to construct a compact MPO, several methods have been proposed. The most commonly used method in quantum chemistry is to design MPO symbolically (or sometimes called analytically) by hand through inspecting the recurrence relation between neighboring sites.~\cite{chan2016matrix} The so-called complementary operator technique is always fully explored to make the MPO more compact, which is essential to the operators with long-range interactions,~\cite{xiang1996density} such as the ab initio electronic Hamiltonian. 
Though usually this method could give the optimal answer by a smart design, it is not automatic in that different operators need a re-design and a re-implementation.
The second one is a numerically ``top-down'' algorithm in which a naive MPO is first constructed and then compressed by the singular value decomposition (SVD) or by removing the linearly dependent terms.~\cite{hubig2017generic} This algorithm is generic and automatic for different operators, while a numerical error is introduced and its effect on the following calculations cannot be well quantified in advance. Apart from this, the time cost spent on the numerical compression is not negligible when the number of terms in the operator is large. 
The third one which is not widely used in quantum chemistry is to construct a finite-state automaton to mimic the interaction terms in the operator.~\cite{crosswhite2008finite} The automaton is easy to be constructed for a translationally invariant lattice model with short-range interactions, but becomes extremely complicated for long-range interactions.

Unlike the ab-initio electronic Hamiltonian which has the same formula for different systems and thus could be hard-coded in implementation,  a general Hamiltonian could be completely different according to the different interactions within the system. Thus, it is not efficient to use the first hand-crafting method mentioned above to construct MPOs on a case-by-case basis. In addition to the inefficiency, it is also difficult to obtain a globally optimal MPO when the Hamiltonian is very complicated.
Therefore, it is necessary and desired to have a better MPO construction algorithm which has all the advantages of the methods introduced above:
\begin{enumerate*}[label=(\roman*)]
\item It is generic for all types of operators with an analytical SOP form;
\item It is automatic, directly from the symbolic operator strings to the MPO;
\item It gives an optimal MPO. Here, ``optimal'' means that the MPO is as compact as possible globally in a given order of DoF; 
\item It is symbolic thus free of any numerical error.
\end{enumerate*}
In this work, we propose a new MPO construction algorithm which meets all the four requirements based on the graph theory for a bipartite graph.
The remaining sections of this paper are arranged as follows. In section \ref{sec:method}, we will present the idea of the new algorithm and the implementation details. In section \ref{sec:results}, several typical Hamiltonians are examined ranging from the simple spin-boson model and Holstein model to the more complicated ab initio electronic Hamiltonian and vibrational Hamiltonian described by a sextic force field. All the calculations are carried out with our in-house code Renormalizer.~\cite{renormalizer} The resulting MPOs are compared with the optimally hand-crafted ones reported in the literature.

\section{Methodology and Implementation}
\label{sec:method}

\subsection{MPO and complementary operator technique}
\label{subsec: mpo}
The wavefunction ansatz in DMRG is called the matrix product states or tensor train, which is 
\begin{gather}
    |\Psi\rangle  = \sum_{\{a\},\{\sigma\}}
     A[1]^{\sigma_1}_{a_1} A[2]^{\sigma_2}_{a_1a_2} \cdots
           A[N]^{\sigma_N}_{a_{N-1}}  | \sigma_1\sigma_2\cdots\sigma_N \rangle \label{eq:mps}.
\end{gather}
For a system of distinguishable particles, $N$ is the number of DoFs in the system and $\{|\sigma_i\rangle\}$ is the local basis such as the discrete variable representation (DVR) basis for nuclear motion. For electronic systems, $N$ is the number of orbitals and $\{|\sigma_i\rangle\}$ is the occupation configuration of each orbital (if using spatial-orbital, $\{|\sigma_i\rangle\} = \{|\textrm{vacuum} \rangle, | \uparrow \rangle, |\downarrow \rangle, |\uparrow \downarrow \rangle\}$; if using spin-orbital, $\{|\sigma_i\rangle\} = \{|\textrm{vacuum} \rangle, | \textrm{occupied} \rangle\}$.).
$\{A[i]^{\sigma_i}_{a_{i-1}a_{i}}\}$ are the local matrices connected by the indices $a_i$, which is commonly called (virtual) bond with bond dimension $M_{\textrm{S}}$ or denoted as $|a_i|$. $\sigma_i$ is called the physical bond with dimension $d$. One good feature of DMRG is that the accuracy is only determined by the dimension of the virtual bond, and thus could be systematically improved.

Similar to MPS, any operator $\hat{O}$ could be expressed as a matrix product operator:~\cite{schollwock2011density,chan2016matrix}
\begin{gather}
    \label{eq:mpo}
    \hat{O} = \sum_{\{w\},\{\sigma\},\{\sigma'\}}
     W[1]^{\sigma'_1, \sigma_1}_{w_1} W[2]^{\sigma'_2, \sigma_2}_{w_1w_2} \cdots
                    W[N]^{\sigma'_N, \sigma_N}_{w_{N-1}} 
                    | \sigma'_1\sigma'_2\cdots\sigma'_N \rangle
                    \langle \sigma_N\sigma_{N-1} \cdots \sigma_1 |.
\end{gather}
MPO could be constructed by sequential singular value decompositions from the matrix element representation $\mathbf{O}_{\sigma_1'\sigma_2'\cdots \sigma_N', \sigma_1\sigma_2\cdots \sigma_N}$ numerically, but it is not practical for a large system since the exact decomposition needs the bond dimension $M_O$ to increase exponentially, which is $d^2, d^4, \cdots, d^{N-2}, d^N, d^{N-2}, \cdots, d^2$ if $N$ is even.  In practice, if an operator has an SOP form, MPO is usually first constructed symbolically, 
\begin{align}
    \hat{O} & =\sum_{\{z\}} \gamma_{z_1z_2\cdots z_N} \hat{z}_1 \hat{z}_2 \cdots \hat{z}_N \label{eq:mpo_operator_basis}  \\
    & =  \sum_{\{w\},\{z\}} W[1]^{z_1}_{w_1}  W[2]^{z_2}_{w_1w_2}  \cdots W[N]^{z_N}_{w_{N-1}} \hat{z}_1 \hat{z}_2 \cdots \hat{z}_N   \label{eq:mpomps}\\
    &  = \sum_{\{w\}}  \hat{W}[1]_{w_1} \hat{W}[2]_{w_1 w_2} \cdots   \hat{W}[N]_{w_{N-1}}.   \label{eq:mpo2}
\end{align}
In Eq.~\eqref{eq:mpo_operator_basis}, $\{\hat{z}_i\}$ represents the elementary operators of each local site such as $\{ \hat{I}, \hat{p}^2, \hat{x}, \hat{x}^2, f(\hat{x},\hat{p}), \textrm{etc}\}$ for a vibrational site or $\{\hat{I}, \hat{a}^\dagger, \hat{a}, \hat{a}^\dagger \hat{a} \}$ for an electronic site. The prefactor $\gamma_{z_1z_2\cdots z_N}$ is commonly very sparse. For example, in the ab initio electronic Hamiltonian, $\gamma_{z_1z_2\cdots z_N}=0$ if more than four $\hat{z}_i$ are $\hat{a}^\dagger$ or $\hat{a}$.
$\gamma_{z_1z_2\cdots z_N}$ could be regarded as the coefficient of $\hat{O}$ on the operator basis $\hat{z}_1 \hat{z}_2 \cdots \hat{z}_N$ and its  matrix product representation in Eq.~\eqref{eq:mpomps} is very similar to an MPS in Eq.~\eqref{eq:mps}.
In Eq.~\eqref{eq:mpo2}, $\hat{W}[i] = \sum_{z_i} W[i]^{z_i} \hat{z}_i$ is a matrix composed of some prefactor attached symbolic operators acting locally on site $i$. 
From this symbolic MPO, it is easy to obtain the matrix element  representation as Eq.~\eqref{eq:mpo} by expanding $\hat{W}[i]$ on the local basis $\{|\sigma_i\rangle\}$. 

From $\gamma_{z_1z_2\cdots z_N}$, if all terms with a nonzero prefactor are extracted, $\hat{O}$ can also be expressed as
\begin{gather}
    \hat{O} = \sum_{o=1}^K \hat{O}[1:N]_o =\sum_{o=1}^K (\gamma_{o} \prod_{i=1}^N \hat{z}_i^o). \label{eq:sop}
\end{gather} 
$K$ is the number of nonzero terms in total. $\hat{z}_i^o$ is the local operator of the $o$th term at site $i$ and could be any of the elementary operators in $\hat{z}_i$. The slice $[1:N]$ indicates that the operator is from site $1$ to site $N$. 
The MPO representation of each term  $\hat{O}[1:N]_o$ in Eq.~\eqref{eq:sop} has $M_O = 1$ with $\hat{W}[i] = \hat{z}_i^o$ and the prefactor $\gamma_o$ could be attached to any site.  The global arithmetic addition of any two MPOs (not necessary to have $M_O = 1$) is
\begin{gather}
    \gamma_1\hat{O}[1:N]_1 + \gamma_2\hat{O}[1:N]_2
    =
    \begin{bmatrix}
    \hat{z}_1^1 &  \hat{z}_1^2 
\end{bmatrix}\big(
\prod_{i=2}^{N-1}
\begin{bmatrix}
   \hat{z}_i^1 &  \mathbf{0}  \\
   \mathbf{0} & \hat{z}_i^2 
\end{bmatrix} \big)
\begin{bmatrix}
   \gamma_1 \hat{z}_N^1 \\
   \gamma_2 \hat{z}_N^2
\end{bmatrix} , \label{eq:mpoadd}
 \end{gather}
which merges the local matrices block-diagonally. Therefore, the naive way to construct MPO of $\hat{O}$ in Eq.~\eqref{eq:sop} will give $M_O = K$. 
 
A more systematic way to derive MPO is to use the recurrence relation between the neighboring sites. When the system is split between site $i$ and site $i+1$ into the respective left (L, from site 1 to $i$) and right (R, from site $i+1$ to $N$) blocks, $\hat{O}$ could be expressed as
\begin{gather}
    \hat{O} = \sum_{o_i=1}^K  \gamma_{o_i} \cdot \hat{O}[1:i]_{o_i} \otimes \hat{O}[i+1:N]_{o_i} \label{eq:split}
\end{gather}
$\hat{O}[1:i]_{o_i} = \prod_{j=1}^i \hat{z}_j^{o_i}$ and $\hat{O}[i+1:N]_{o_i}=  \prod_{j=i+1}^N \hat{z}_j^{o_i}$  are usually called the normal operators.  A recurrence relation between the neighboring $\hat{O}[1:i-1]_{o_{i-1}}$ and $\hat{O}[1:i]_{o_{i}}$ could be defined as
\begin{gather}
   \hat{O}[1:i]_{o_{i}} =  \sum_{o_{i-1}=1}^K \hat{O}[1:i-1]_{o_{i-1}}  \hat{O}[i]_{o_{i-1}o_{i}},
\end{gather}
from which the symbolic MPO in Eq.~\eqref{eq:mpo2} could be obtained directly with $\hat{W}[i] = \hat{O}[i]$ and again the prefactor $\gamma_{o_i}$ could be attached to any site. 
This construction gives the same result as the global arithmetic addition of $K$ MPOs with $M_O=1$ in Eq.~\eqref{eq:mpoadd}. However, it is apparently not optimal in that some of the interaction terms in Eq.~\eqref{eq:split} may share the common operators in the set $\{\hat{O}[1:i]_{o_i}\}$ or $\{\hat{O}[i+1:N]_{o_i}\}$. For example, if $K=2$ and $\hat{O}[i+1:N]_{1} \equiv \hat{O}[i+1:N]_{2}$ while $\hat{O}[1:i]_{1} \neq \hat{O}[1:i]_{2}$, $\hat{O}[1:i]_1$ and $\hat{O}[1:i]_{2}$ could be summed up with the prefactors to create a complementary operator $\hat{\tilde{O}}[1:i]_{1} = \gamma_1 \hat{O}[1:i]_1 + \gamma_2 \hat{O}[1:i]_{2}$ on the L-block and meanwhile $\hat{O}[i+1:N]_{2}$ is removed from the R-block so that $\hat{O} = \hat{\tilde{O}}[1:i]_{1} \otimes \hat{O}[i+1:N]_{1}$. Thus, $|o_i|$, the number of columns of $\hat{O}[i]$, is reduced by 1. This example shows that the MPO representation of the same operator is not unique as long as the product result is correct.
Generally speaking, to make the MPO compact, if there are redundant operators in $\{\hat{O}[i+1:N]_{o_i}\}$($\{\hat{O}[1:i]_{o_i}\}$), the corresponding left (right) complementary operators could be created. This complementary operator technique~\cite{xiang1996density} is of essential importance in constructing MPO for ab initio electronic Hamiltonian by assembling all the 4-index operators $\sum_{pqrs} g_{pqrs} a^\dagger_p a^\dagger_q a_r a_s$ and 3-index operators $\sum_{pqr} g_{pqrs} a^\dagger_p a^\dagger_q a_r $   and part of the 2-index operators in one block, reducing $M_O$ from $\mathcal{O}(N^4)$ to $\mathcal{O}(N^2)$.~\cite{keller2015efficient, chan2016matrix} 
However, the complexity of designing complementary operators comes from that in most Hamiltonian both $\hat{O}[1:i]_{o_i}$ and $\hat{O}[i+1:N]_{o_i}$ in one interaction term are correlated to other interaction terms. For instance, we add another two terms in the former example, $\hat{O}[1:i]_{1} \equiv \hat{O}[1:i]_{3}$ and $\hat{O}[1:i]_{2} \equiv \hat{O}[1:i]_{4}$. In this case, the optimal solution is to create complementary operators  $\hat{\tilde{O}}[i+1:N]_{1} = \gamma_1 \hat{O}[i+1:N]_{1} + \gamma_3 \hat{O}[i+1:N]_{3}$ and $\hat{\tilde{O}}[i+1:N]_{2} = \gamma_2 \hat{O}[i+1:N]_{2} + \gamma_4 \hat{O}[i+1:N]_{4}$, which will give $|o_i|=2$. While creating the complementary operator $\hat{\tilde{O}}[1:i]_{1}$ as above will result in $|o_i|=3$. This toy example shows that the design of complementary operators is nontrivial.  A typical real example is
that when constructing MPO of the ab initio electronic Hamiltonian, a different design strategy of the complementary operators of the 2-index operators within one block will lead to a different $M_O$ shown in Figure 10 of Ref.~\citenum{chan2016matrix}  though all of them are $\mathcal{O}(N^2)$. 
Therefore, the key to construct a compact MPO is to design and select the normal and complementary operators smartly at each bond to make the number of retained operators as small as possible. As far as we know, up to now it is still an art to design the complementary operators by hand on a case-by-case basis rather than by a rigorous and automatic procedure.

\subsection{MPO construction algorithm via bipartite graph theory}
\label{sec:core_alg}
We propose to use the theory of bipartite graph to set a rigorous foundation to construct MPO automatically.
We first reinterpret the operator selection problem at each bond mentioned in section \ref{subsec: mpo} as a minimum vertex cover problem in a bipartite graph and then prove that the locally optimal solution is also globally optimal. 

\begin{figure}[htbp]
\centering
\includegraphics[width = 0.25 \textwidth]{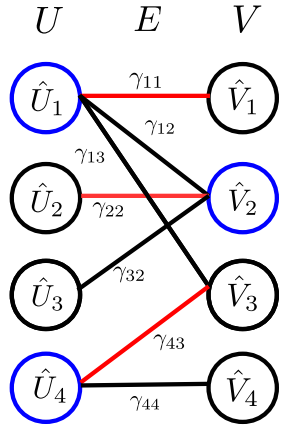}  

\caption{\label{fig:bipartite_graph}
An example of mapping the operator $\hat{O} = \gamma_{11} \hat{U}_1 \hat{V}_1 + \gamma_{12}\hat{U}_1 \hat{V}_2  + \gamma_{13}\hat{U}_1 \hat{V}_3 +  \gamma_{22}\hat{U}_2 \hat{V}_2 + \gamma_{32}\hat{U}_3 \hat{V}_2 + \gamma_{43}\hat{U}_4 \hat{V}_3 + \gamma_{44}\hat{U}_4 \hat{V}_4$ to a bipartite graph $G=(U, V, E)$. The vertices represent the non-redundant operators in the L- and R- block. The edges represent the  interactions with a nonzero prefactor. The vertices in blue form a minimum vertex cover. The edges in red form a maximum matching. 
}
\end{figure}

The non-redundant operator set by removing the duplicated operators in $\{\hat{O}[1:i]_{o_i}\},\{\hat{O}[i+1:N]_{o_i}\}$ of Eq.~\eqref{eq:split} are denoted as $U = \{\hat{U}[1:i]_{u_i}\}, V = \{\hat{V}[i+1:N]_{v_i}\}$, which are represented as the vertices in Fig.~\ref{fig:bipartite_graph}. Unlike that the interaction pattern is one-to-one between $\{\hat{O}[1:i]_{o_i}\}$ and $\{\hat{O}[i+1:N]_{o_i}\}$, it would be one-to-many between $\{\hat{U}[1:i]_{u_i}\}$ and $\{\hat{V}[i+1:N]_{v_i}\}$. The $K$ interaction terms are represented as the edges denoted as $E$ each connecting one vertex in $U$ to one vertex in $V$ with a prefactor (weight) $\gamma_{u_i v_i}$. A bipartite graph is often denoted as $G=(U,V,E)$. If the $p$th vertex in $U$ is selected, the corresponding operator $\hat{U}[1:i]_p$ in the L-block is retained. 
Meanwhile, the operators $\hat{V}[i+1:N]_q$ corresponding to the vertices in $V$ which are linked to $\hat{U}[1:i]_p$ through edges are multiplied by the prefactor of the certain edge and then are added up to create a new complementary operator in the R-block $\sum_q \gamma_{pq} \hat{V}[i+1:N]_q $. 
The same rule is applied if a vertex in $V$ is selected.
Therefore, the minimal number of retained operators in one block which could cover all the $K$ interaction terms is equal to the minimal number of selected vertices in $(U, V)$ which could cover all the edges in $E$ (shown in blue in Fig.~\ref{fig:bipartite_graph}). 
The latter problem is called the minimum vertex cover in graph theory. For a bipartite graph described here, K{\"o}nig theorem proves that the number of vertices in the minimum vertex cover is equal to the number of edges in the maximum matching.~\cite{bondy1976graph} A matching is an edge set in which any two edges do not share one vertex. The maximum matching shown in red in Fig.~\ref{fig:bipartite_graph} is the matching having the maximal number of edges, which could  be solved efficiently by the Hungarian algorithm~\cite{kuhn1955hungarian} with complexity $\mathcal{O}(nm)$ or the Hopcroft–Karp algorithm~\cite{hopcroft1973n} with complexity $\mathcal{O}(\sqrt{n}m)$ through finding an augmenting path.~\cite{bondy1976graph} Here, $n$ and $m$ are the total number of vertices and edges in the bipartite graph. Once the maximum matching is found, the vertices in the minimum vertex cover could be obtained easily and the retained operators are optimally selected according to the rules above. 

For a DMRG chain with a certain order, the whole procedure to construct the MPO of $\hat{O}$ from site $1$ to $N$ (from $N$ to $1$ is similar) is summarized as follows. 

\begin{enumerate}[label*=\arabic*.]
\item  \label{item:step1} The incoming non-redundant operator set of site $i$ is known as $\{ \hat{W}[1:i-1]_{w_{i-1}} \}$ ($\{\hat{W}[1:0]\}$ is $\{1\}$), which are also the outgoing operators of site $i-1$. 
Commonly, $\{ \hat{W}[1:i-1]_{w_{i-1}} \}$ includes both normal operators and complementary operators.
Next, $\{ \hat{W}[1:i-1]_{w_{i-1}} \}$ is multiplied by the local elementary operators $\{\hat{z}_i\}$ on site $i$ to form a non-redundant operator set $\{\hat{U}[1:i]_{u_i}\} = \{\hat{W}[1:i-1]_{w_{i-1}}\} \otimes \{\hat{z}_i\}$. The R-block non-redundant operator set is $\{\hat{V}[i+1:N]_{v_i}\}$, in which all operators are normal operators. 
Note that for efficiency only the interaction terms with a nonzero prefactor are necessary to be included in the operator sets $\{\hat{U}[1:i]_{u_i}\}$ and $\{\hat{V}[i+1:N]_{v_i}\}$. Hence, 
at this boundary between site $i$ and $i+1$, $\hat{O} = \sum_{u_i v_i} \gamma_{u_i v_i} \hat{U}[1:i]_{u_i} \otimes \hat{V}[i+1:N]_{v_i} $.

\item  \label{item:step2}  The operators in $\{\hat{U}[1:i]_{u_i}\}$, $\{\hat{V}[i+1:N]_{v_i}\}$ and the interactions between them are represented as vertices and edges to form a bipartite graph $G = (U, V, E)$ (see Fig.~\ref{fig:bipartite_graph}). Afterward, the maximum matching and the corresponding minimum vertex cover of this bipartite graph is found with the Hungarian algorithm or the Hopcroft–Karp algorithm. Next, iterating through each vertex in the minimum vertex cover once:
\begin{enumerate}[label*=\arabic*.]
\item If the vertex is the $p$th vertex in $U$, the operator $\hat{U}[1:i]_p$ is retained and meanwhile the edges linked to it are removed from the graph. 
\item  If the vertex is the $q$th vertex in $V$, the complementary operator linked through edges to $\hat{V}[i+1:N]_q$ is created and retained, which is $\hat{\tilde{U}}[1:i]_q = \sum_{p} \gamma_{pq} \hat{U}[1:i]_p $. Meanwhile, the edges are removed. 
\end{enumerate}
 The reason to remove the edges after each visit is to avoid the double-counting of the interactions. After all the vertices in the minimum vertex cover are visited once, there will be no edge in the graph.
 
\item \label{item:step3} The retained operators $\hat{U}[1:i]_p$ and $\hat{\tilde{U}}[1:i]_q$ together form a new non-redundant operator set $\{\hat{W}[1:i]_{w_{i}} \}$ in the L-block. It is the outgoing operator set of site $i$ and meanwhile is the incoming operator set of site $i+1$. 
After that, with $\{\hat{
W}[1:i-1]_{w_{i-1}}\}$  and $\{\hat{W}[1:i]_{w_i}\} $, the local symbolic MPO $\hat{W}[i]$ is easy to obtain according to the recurrence relation $\hat{W}[1:i] =\hat{W}[1:i-1] \hat{W}[i]$.
In fact, the local prefactor matrix $W[i]^{z_i}_{w_{i-1}w_i}$ in $\hat{W}[i]_{w_{i-1}w_{i}} = \sum_{z_i} W[i]^{z_i}_{w_{i-1}w_i} \hat{z}_i$ is 
the transformation matrix (reshaped to be $W[i]_{w_{i-1}z_i, w_i}$) of operator basis from $\{\hat{W}[1:i-1]_{w_{i-1}}\} \otimes  \{\hat{z}_i\}$ to $\{\hat{W}[1:i]_{w_{i}} \}$.

Return back to step \ref{item:step1}.
\end{enumerate}

The procedure described above is apparently a locally optimal solution, since the selected operators have already been the minimum vertex cover at each boundary when sweeping from the left to the right. 
To prove that the locally optimal solution is also globally optimal, we should prove that at each boundary between site $i$ and $i+1$, the number of edges in the maximum matching (the number of vertices in the minimum vertex cover) is the same no matter whether the operator set of L-block is composed of all normal operators or is composed of both normal operators and complementary operator as  $\{\hat{W}[1:i]_{w_{i}} \}$ according to step \ref{item:step1} to step \ref{item:step3}.
Following Eq.~\eqref{eq:mpo_operator_basis}, if the coefficient tensor $\gamma_{z_1z_2 \cdots z_N}$ is reshaped as a matrix $\gamma_i = \gamma_{z_1z_2 \cdots z_i, z_{i+1} \cdots z_N}$, it could be regarded as the coefficient matrix of $\hat{O}$ expanded on the operator basis $\{\hat{z}_1 \otimes \cdots \otimes \hat{z}_i\} \otimes \{\hat{z}_{i+1} \otimes\cdots \otimes\hat{z}_N\} $ in the operator space. $\gamma_i$ is called the unfolding matrix of $\gamma$ in Ref.~\citenum{oseledets2011tensor}, whose rank is denoted as $r_i$ called TT-rank. 
The bipartite graph $G[i]= (U[i],V[i],E[i])$ at the boundary between site $i$ and site $i+1$ is $U[i] = \{\hat{z}_1 \otimes \cdots \otimes \hat{z}_i\}$, $V[i]=\{\hat{z}_{i+1} \otimes\cdots \otimes\hat{z}_N\}$, the edges $E[i]$ has a one-to-one correspondence to the nonzero matrix elements in $\gamma_i$.
In the bipartite graph theory,  the matrix $\gamma_i$ could also be regarded as a symbolic bipartite adjacency matrix, for which only that the matrix elements are zero or nonzero is important. Lov\'asz proposed the theorem that the rank of the symbolic adjacency matrix is equal to the  number of edges of a maximum matching.~\cite{lovasz1979determinants} Therefore, since $U[i]$ and $V[i]$ are composed of all normal operators, using the rules described above to select the normal and complementary operators, the ideally minimal number of retained operators at this boundary is equal to  $r_i$, the rank of matrix $\gamma_i$. \add{In Appendix \ref{sec:global_opt}, we prove that sweeping from left to right as the procedure above will not change the rank of the  adjacency matrix at the same boundary.}
It is worth noting that in Ref.~\citenum{hubig2017generic}, the ideal rank $r_i$ of MPO at the $i$th bond is expected to be approached by numerical SVD compression, deparallelization and delinearization, but it is not guaranteed because of the numerical error. But here, it is guaranteed symbolically via the bipartite graph theory. \add{In addition, the scaling of the current algorithm is roughly $\mathcal{O}(K^{3/2}N)$ with Hopcroft–Karp algorithm. In comparison, The scaling of the SVD-based algorithm is roughly $\mathcal{O}(K^3d^2N)$. Thus, the current algorithm is much cheaper.}

Several other advantages of the algorithm are that 
\begin{enumerate*}[label=(\roman*)]
\item The sparsity of MPO is fully maintained, which could be used to reduce the computational cost during the tensor contraction in DMRG single state or time evolution algorithms.
\item The symmetry could be directly implemented by attaching the good quantum numbers on each normal and complementary operator.
\item  The algorithm not only works for MPO construction, but also works for MPS construction if the wavefunction in the Fock space representation has already been known. For the same reason, the obtained MPS is the most compact one to represent the wavefunction exactly.  
\end{enumerate*}
    
Finally, it should be mentioned that for a system in which the interaction pattern is inhomogeneous, the order of DoFs will affect the size of MPO. It is still unclear whether there is an algorithm which could efficiently find out a specific order giving the minimal MPO. However, in our opinion, this problem is less of a priority than the widely known ordering problem with respect to the accuracy of DMRG calculation.~\cite{legeza2003optimizing,moritz2005convergence}

\section{Results}
\label{sec:results}
In this section, we will demonstrate the effectiveness of the new algorithm by constructing the MPOs of Hamiltonians ranging from the simple spin-boson model and Holstein model to the more complicated ab initio electronic Hamiltonian and vibrational Hamiltonian with a sextic force field.

\subsection{Spin-boson model and Holstein model}

\begin{figure*}[htbp]
\centering 

\subfloat[]{
    \label{fig:sbm}
    \includegraphics[width = 0.33 \textwidth]{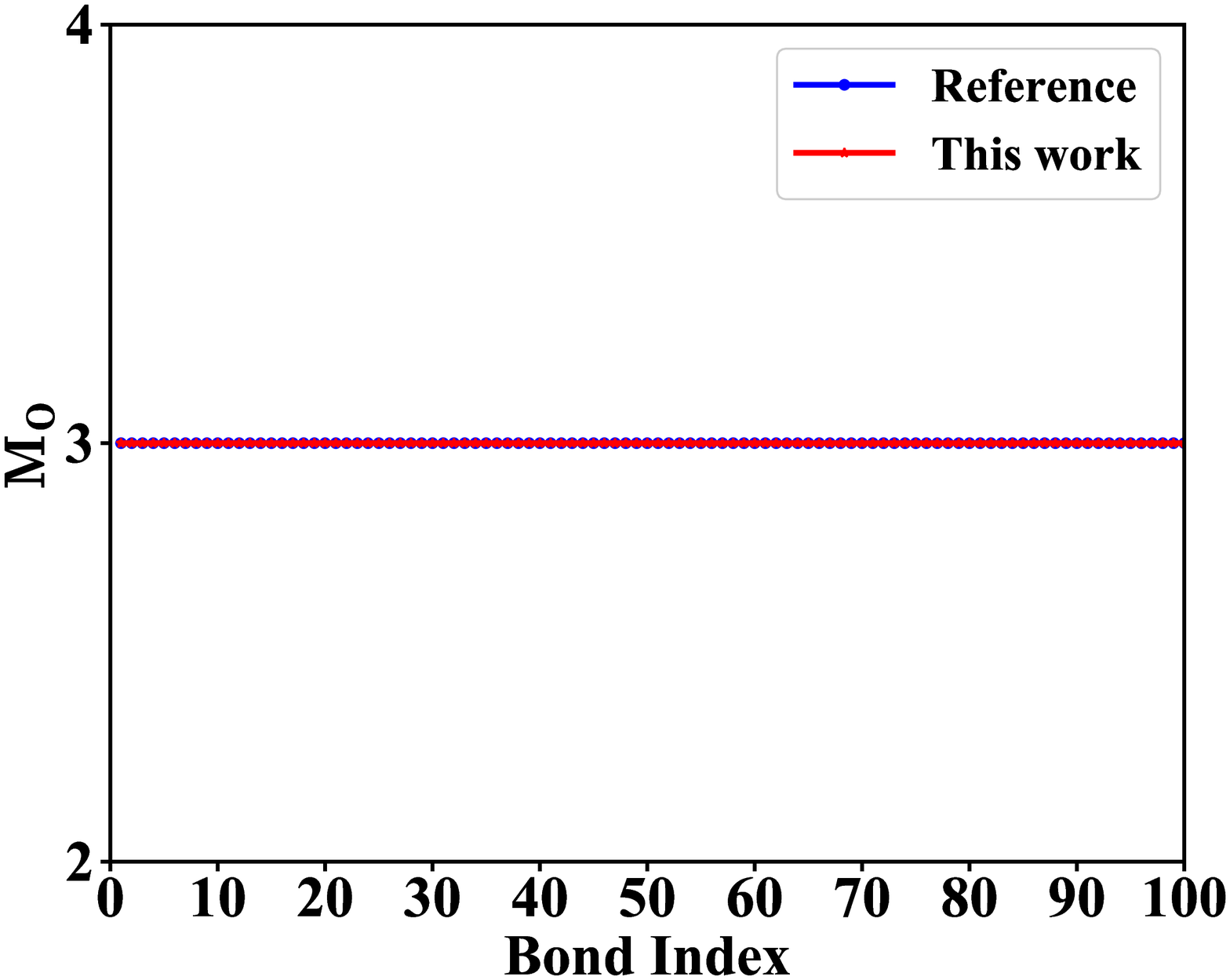}  
    } 
\subfloat[]{
    \label{fig:holstein20}
    \includegraphics[width = 0.33 \textwidth]{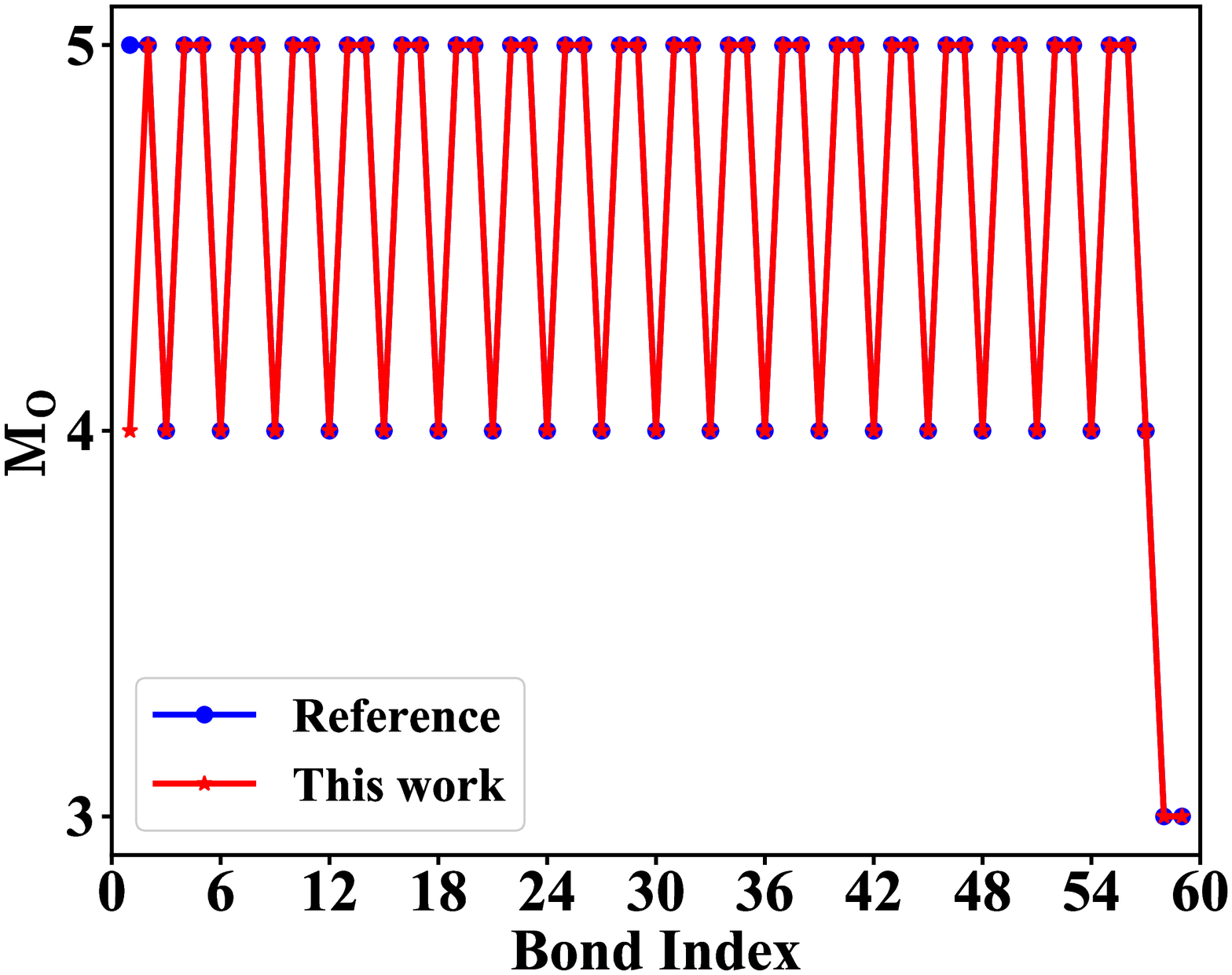}
    }
\subfloat[]{
    \label{fig:holstein20_longrange}
    \includegraphics[width = 0.33 \textwidth]{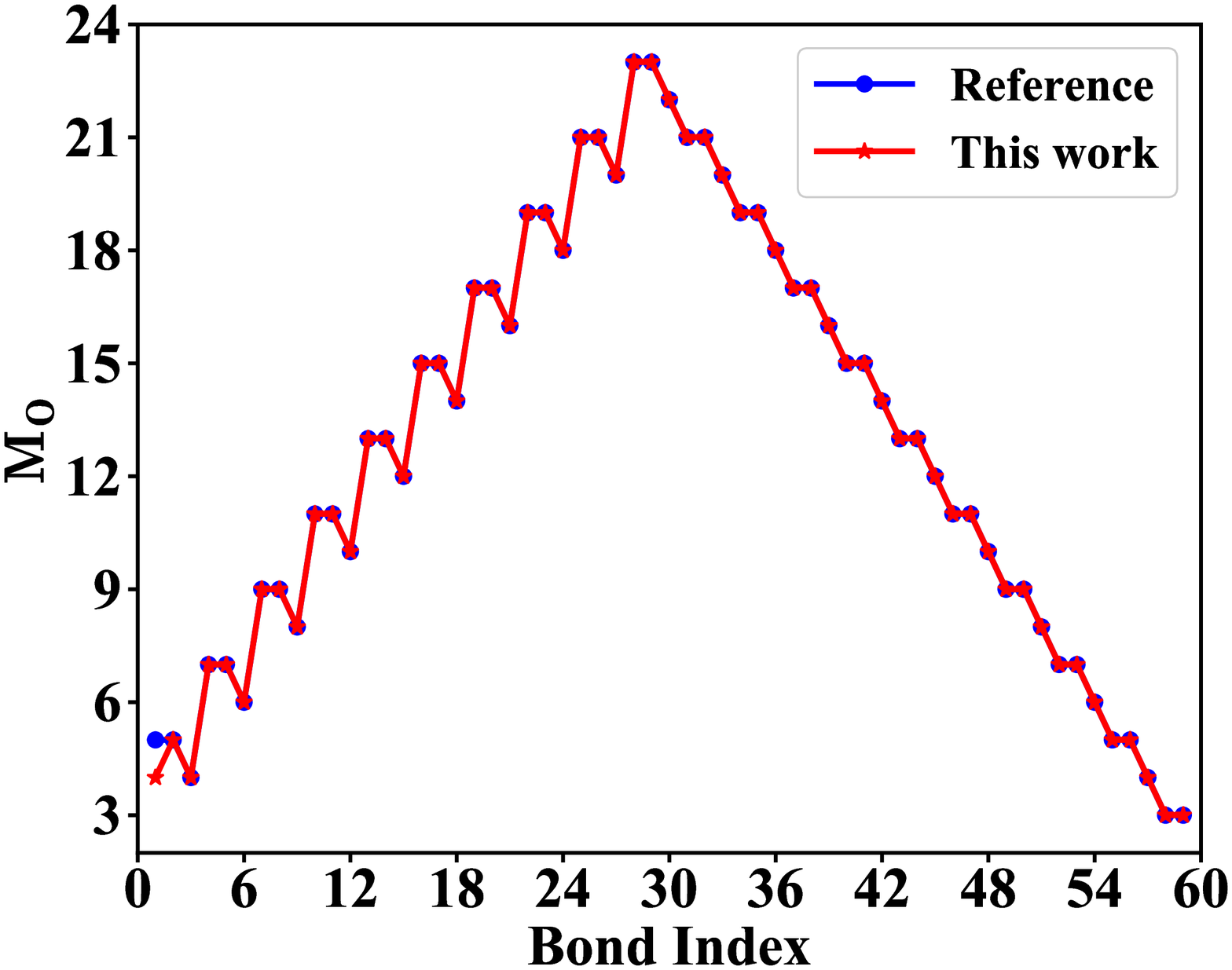}
    }
    
\caption{\label{fig:sbm_holstein}
The bond dimension $M_O$ versus the bond index in (a) spin-boson model with 100 discrete vibrational modes. (b) Holstein model of 20 electronic sites with only one-dimensional electronic coupling and each electronic site has two vibrational modes. (c) same as (b) except with arbitrary long-range electronic couplings. The reference results (blue line) are based on the hand-crafted complementary operator strategy provided in our former work.~\cite{ren2018time}
}
\end{figure*}

The spin-boson model (expressed in the first quantization formalism in Eq.~\eqref{eq:sbm}) describes a two-level system coupled with a harmonic bath, which is widely used to investigate the quantum dissipation. 
\begin{gather}
    \hat{H}_{\textrm{SBM}} = \epsilon \hat{\sigma}_z + \Delta \hat{\sigma}_x + \frac{1}{2}
    \sum_i(\hat{p}_i^2+\omega^2_i \hat{q}_i^2) + \hat{\sigma}_z \sum_i c_i \hat{q}_i \label{eq:sbm}
\end{gather}
Holstein model (expressed in the second quantization formalism in Eq.~\eqref{eq:Holstein}) is also a widely used electron-vibrational coupling model to describe the charge transport,  energy transfer and spectroscopy of molecular aggregates.~\cite{schroter2015exciton,ren2018time, li2020numerical,Jiang2020finite,Li2020finite} It could be regarded as a group of two-level systems as the spin-boson model coupled with each other through coupling constant $J_{ij}$. 
\begin{gather}
        \hat{H}_{\textrm{Holstein}} = \sum_i \varepsilon_i a_i^\dagger a_i + 
        \sum_{i\neq j} J_{ij} a_i^\dagger a_j 
        + \sum_{in} \omega_{in} b_{in}^\dagger b_{in}
        + \sum_{in} \omega_{in} g_{in} a_i^\dagger a_i (b_{in}^\dagger + b_{in}) 
        \label{eq:Holstein}
\end{gather}
Both of the two models are often adopted to benchmark the quantum dynamics methods. We put the two models in the same section because spin-boson model could be regarded as a one-site Holstein model with an additional interstate coupling $\Delta$ and thus the MPOs of them are very similar.
We test a spin-boson model with 100 discrete modes and the order is $[\textrm{spin}, v_1, v_2 \cdots ,v_{100}]$. 
We also test two Holstein models with 20 electronic sites and both of them have two vibrational modes of each electronic site but the former only has one-dimensional nearest-neighbor electronic hopping while the latter has long-range hoppings between any two electronic sites.
The order of the Holstein model is 
$[e_1, v_{1,1}, v_{1,2}, e_2, v_{2,1}, v_{2,2}, \cdots, e_{20}, v_{20,1}, v_{20,2}]$.
The MPO bond dimension $M_O$ versus the bond index is shown in Fig.~\ref{fig:sbm_holstein}. The reference results (blue line) are based on a hand-crafted strategy, in which the normal operators for the electronic coupling terms are switched to the complementary operators $\hat{P}_j = \sum_i J_{ij} a_i$ and $\hat{P}_j^\dagger = \sum_i J_{ij} a^\dagger_i$ after passing the middle electronic site. The details are provided in the appendix in our former work,~\cite{ren2018time} which is believed to be near-optimal for the two models (from the results shown below, it is optimal except at the first bond for the Holstein model.). 

For the spin-boson model shown in Fig.~\ref{fig:sbm}, $M_O$ is a constant independent of system size because $\hat{W}[1:i] = \{ \hat{H}[1:i], \hat{\sigma}_z, \hat{I} \}$ where $\hat{H}[1:i]$ is the complete Hamiltonian from site 1 to $i$. The new automatic algorithm gives exactly 
the same result as the hand-crafted one. For the Holstein model shown in Fig.~\ref{fig:holstein20} and ~\ref{fig:holstein20_longrange}, $M_O$ is independent of the number of the electronic site when the electronic coupling is one-dimensional nearest-neighbor coupling, while it is linearly dependent on the number of the electronic site if the long-range hopping is allowed. 
The new automatic algorithm gives the same results as the hand-crafted ones except at the first bond, where the new algorithm gives one less bond dimension.
This minor difference comes from that the hand-crafted strategy gives $\hat{W}[1] = \{\varepsilon_1 a_1^\dagger a_1, a^\dagger_1 a_1, a^\dagger_1, a_1, \hat{I} \}$ while the automatic algorithm gives $\hat{W}[1] = \{a^\dagger_1 a_1, a^\dagger_1, a_1, \hat{I} \}$ and the local energy of the first site $\varepsilon_1 a^\dagger_1 a$ is considered in $\hat{W}[2]_{0,0} = \varepsilon_1 \hat{I}$. Though this small improvement will not make a noticeable difference on the actual computational cost, it is clear to demonstrate that since the new algorithm is globally optimal, it could find out the redundancy which will be neglected sometimes with the common hand-crafted strategy. 

\subsection{Ab initio electronic Hamiltonian}
\label{sec:abH}
The second Hamiltonian considered is the ab initio electronic Hamiltonian, in which up to 4 sites interact with each other. Thus it is much more complicated than the spin-boson model and Holstein model.  With spin-orbitals, the Hamiltonian is written as 
\begin{gather}
    \hat{H}_{\textrm{el}} = \sum_{p,q=1}^{N} h_{pq} a_p^\dagger a_q + \frac{1}{2} \sum_{p,q,r,s=1}^{N} v_{pqrs} a_p^\dagger a_q^\dagger a_r a_s = \sum_{p,q=1}^{N} h_{pq} a_p^\dagger a_q + \sum_{p < q,r<s}^{N} g_{pqrs} a_p^\dagger a_q^\dagger a_r a_s
    \label{eq:abinito}
\end{gather}
where the two-electron integral $v_{pqrs}$ is $(ps|qr)$ in chemist's notation. The second equality takes advantage of the symmetry in $v_{pqrs}$ ($g_{pqrs}=v_{pqrs}-v_{qprs} = v_{pqrs}-v_{pqsr}$).

Firstly, we introduce the optimal hand-crafted strategy to construct the MPO of ab initio electronic Hamiltonian. For more implementation details, please refer to Ref.~\citenum{chan2016matrix}. For convenience, $\hat{H}_{\textrm{el}}$ is divided into three components. The first part is $\hat{H}_1 =  \hat{H}_{L}+\hat{H}_{R}$, in which $\hat{H}_{L}$ and $\hat{H}_{R}$ are respectively the full Hamiltonian of the orbitals in the L-block and R-block. In fact, $\hat{H}_{L}$ and $\hat{H}_{R}$ could be regarded as the complementary operators of identity operator $\hat{I}_R$ and $\hat{I}_L$ in the R-block and L-block, reducing $\mathcal{O}(N^4)$ normal operators to 1 complementary operator. Apparently, $\hat{H}_1$ gives $M_{O,1}=2$ at each bond. The second part with two fermionic creation or annihilation (elementary) operators in each block is written as 
\begin{align}
\hat{H}_2 = \sum_{p<q,r<s} & -g_{p_{L} q_{R} r_{L} s_{R}}\left(a_{p_{L}}^{\dagger} a_{r_{L}}\right)\left(a_{q_{R}}^{\dagger} a_{s_{R}}\right) \nonumber \\
  & +g_{p_{L} q_{L} r_{R} s_{R}}\left(a_{p_{L}}^{\dagger} a_{q_{L}}^{\dagger}\right)\big(a_{r_{R}} a_{s_{R}}\big)  \nonumber \\
 & +g_{p_{R} q_{R} r_{L} s_{L}}\big(a_{r_{L}} a_{s_{L}}\big)\left(a_{p_{R}}^{\dagger} a_{q_{R}}^{\dagger}\right)  \label{eq:h2}
\end{align}
The optimal strategy to design the complementary operator depends on the number of orbitals denoted as $n_L$ and $n_R$ in each block. For instance, if $n_L > n_R$, the complementary operators of the first term in Eq.~\eqref{eq:h2} is $\hat{P}_{qs} = \sum_{pr} -g_{p_{L} q_{R} r_{L} s_{R}}\left(a_{p_{L}}^{\dagger} a_{r_{L}}\right)$, which have $n_R^2$ terms in total. Therefore, the ideally minimal bond dimension is
$M_{O,2} = \min(n_L^2, n_R^2) + 2\cdot \min(n_L(n_L-1)/2, n_R(n_R-1)/2)$.
The third part with one  creation or annihilation operator in one block and three in the other is commonly written as 
\begin{align}
\hat{H}_3 = & \sum_{p}    a_{p_{L}}^{\dagger}   \left( \sum_{q} \frac{1}{2} h_{p_L q_R} a_{q_R} + \sum_{qrs} g_{p_{L} q_{R} r_{R} s_{R}} a_{q_{R}}^{\dagger} a_{r_{R}} a_{s_{R}}  \right) \nonumber \\
& + \sum_{r} a_{r_{L}}\left( \sum_{s} - \frac{1}{2} h_{s_R r_L } a_{s_R}^\dagger + \sum_{pqs}  g_{p_{R} q_{R} r_{L} s_{R}}a_{p_{R}}^{\dagger} a_{q_{R}}^{\dagger} a_{s_{R}}\right)  \nonumber \\
& + \sum_q \left(\sum_p -\frac{1}{2} h_{q_R p_L } a_{p_L} + \sum_{psr}  g_{p_{L} q_{R} r_{L} s_{L}} a_{p_L}^{\dagger} a_{r_{L}} a_{s_{L}}\right) a_{q_{R}}^{\dagger} \nonumber \\
& +\sum_s \left( \sum_r \frac{1}{2} h_{r_L s_R} a_{r_L}^\dagger + \sum_{pqr} g_{p_{L} q_{L} r_{L} s_{R}} a_{p_{L}}^{\dagger} a_{q_{L}}^{\dagger} a_{r_{L}}\right) a_{s_{R}}  \label{eq:h3_1}
\end{align}
The terms in the parentheses are the complementary operators which should be firstly summed up. This kind of complementary operators is adopted to construct MPO of ab initio electronic Hamiltonian, because it greatly reduces $M_{O,3}$ from $\mathcal{O}(N^3)$ to $\mathcal{O}(N)$.
However, it is only near-optimal because near the left boundary of the chain, there are more 1-index operators in the R-block than 3-index operators in the L-block. Thus, the optimal way to construct the complementary operator at this boundary is
\begin{align}
\hat{H}_3 = & \sum_{p}    a_{p_{L}}^{\dagger}   \left( \sum_{q}  h_{p_L q_R} a_{q_R} + \sum_{qrs} g_{p_{L} q_{R} r_{R} s_{R}} a_{q_{R}}^{\dagger} a_{r_{R}} a_{s_{R}}  \right) \nonumber \\
& + \sum_{r} a_{r_{L}}\left( \sum_{s} - h_{s_R r_L } a_{s_R}^\dagger + \sum_{pqs}  g_{p_{R} q_{R} r_{L} s_{R}}a_{p_{R}}^{\dagger} a_{q_{R}}^{\dagger} a_{s_{R}}\right)  \nonumber \\
& + \sum_{prs} a_{p_L}^{\dagger} a_{r_{L}} a_{s_{L}} \left( \sum_q g_{p_{L} q_{R} r_{L} s_{L}}  a_{q_{R}}^{\dagger} \right) \nonumber \\
& +\sum_{pqr} a_{p_{L}}^{\dagger} a_{q_{L}}^{\dagger} a_{r_{L}} \left(\sum_s g_{p_{L} q_{L} r_{L} s_{R}}a_{s_{R}}  \right)  \label{eq:h3_2}
\end{align}
The case is the same near the right boundary of the chain.
Therefore, the minimal $M_{O,3}$ equals $2\cdot \min(n_L^2(n_L-1)/2, n_R) + 2 \cdot \min(n_L, n_R^2(n_R-1)/2)$.
It is clear that $M_{O,2}$ contributes most to the total $M_{O} = M_{O,1} + M_{O,2} +M_{O,3}$, and thus this improvement of $M_{O,3}$ is rarely considered. But it could be considered automatically with our new algorithm.
Adding up the contributions of the three components, the largest bond dimension always lies in the middle of the chain, which is  $M_{O, \textrm{max}} = 2(\frac{N}{2})^2 +3(\frac{N}{2}) +2$. 

To consider the antisymmetry of fermions in the algorithm described in section ~\ref{sec:core_alg}, the Jordan-Wigner transformation~\cite{jordan1928pauli} for the elementary creation and annihilation operators is introduced.~\cite{keller2015efficient,li2017spin}
\begin{gather}
    | \textrm{vacuum}\rangle  = | \alpha \rangle \\
    | \textrm{occupied}\rangle  = | \beta \rangle \\
    a^\dagger_j  = \prod_{i=1}^{j-1} \sigma_z[i] \times \sigma_-[j] \\
    a_j  = \prod_{i=1}^{j-1} \sigma_z[i] \times \sigma_+[j] 
\end{gather}
\begin{figure*}[htbp]
\centering
\subfloat[]{
    \label{fig:momax}
    \includegraphics[width = 0.47 \textwidth]{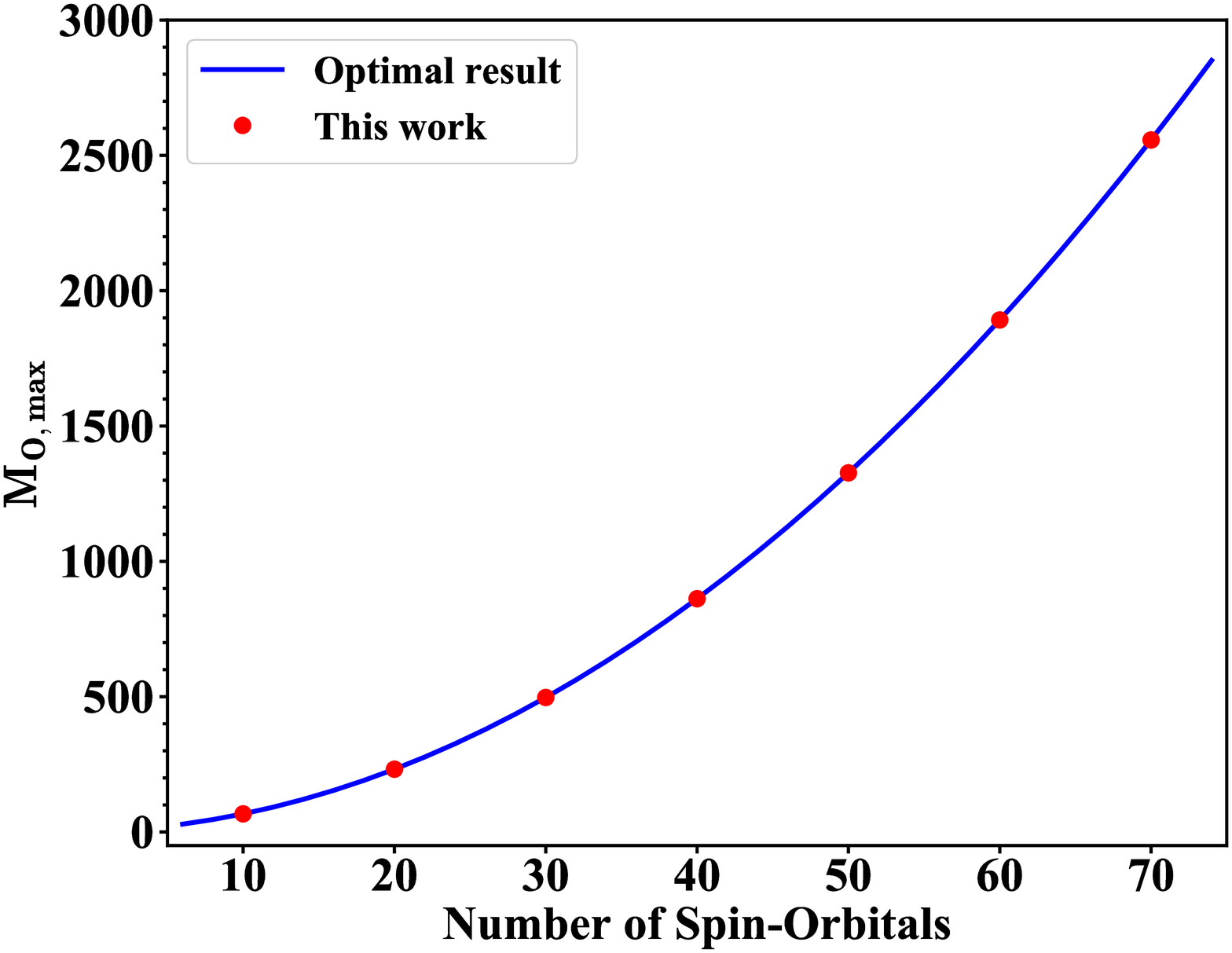}
    }\\
\subfloat[]{
    \label{fig:mo50}
    \includegraphics[width = 0.47 \textwidth]{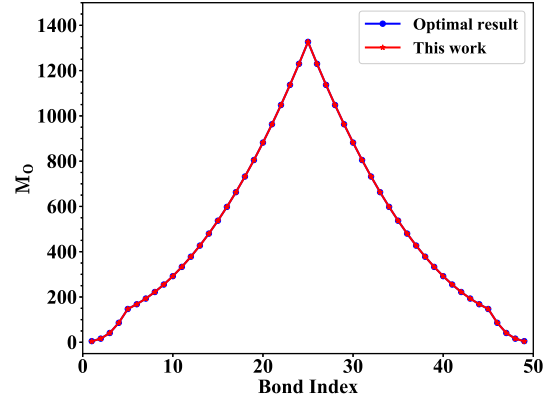}  
    } 
\caption{
(a) The maximal MPO bond dimension $M_{O, \textrm{max}}$ of ab initio electronic Hamiltonian with different number of spin-orbitals. The blue curve $M_{O, \textrm{max}} = 2(\frac{N}{2})^2 +3(\frac{N}{2}) +2$ is the optimal result from the hand-crafted complementary operator strategy (see text for details). The red circles are the results obtained from the new automatic MPO construction algorithm.
(b) The MPO bond dimension $M_O$ at each bond of ab initio electronic Hamiltonian of a 50 spin-orbitals system.
}
\end{figure*}
Fig.~\ref{fig:momax} shows the maximal $M_O$ of systems with 10 to 70 spin-orbitals and Fig.~\ref{fig:mo50} shows $M_O$ at each bond of a system with 50 spin-orbitals. The correctness of the MPOs generated by the automatic algorithm has been verified by checking the residue $\|\textrm{MPO}_1 - \textrm{MPO}_2 \| = 0$ with respect to the MPOs developed by Li \textit{et al.} in Ref.~\citenum{li2017spin} and implemented in package QCMPO.~\cite{QCMPO} $M_{O}$ at each bond and $M_{O, \textrm{max}}$ (red asterisks) of the
automatically generated MPO exactly match what the optimally hand-crafted strategy described above would give (blue circles), except that at the first bond the automatic algorithm gives  $M_O = 4$ ($\hat{W}[1] = [\sigma_- \sigma_+, \sigma_z \sigma_-, \sigma_z \sigma_+, \hat{I}]$)  while the hand-crafted strategy gives $M_O = 5$ ($\hat{W}[1] = [h_{11} \sigma_- \sigma_+, \sigma_- \sigma_+, \sigma_z \sigma_-, \sigma_z \sigma_+, \hat{I}]$). The reason is the same as that in the case of Holstein models.
In addition, in Fig.~\ref{fig:mo50}, $M_{O}$ versus the bond index is symmetric as expected and the kink at the bond index 5 and 45 is due to the switch of the complementary operators from Eq.~\eqref{eq:h3_2} to Eq.~\eqref{eq:h3_1}, indicating that the new algorithm could really find out the optimal solution. 

\begin{figure}[htbp]
\centering
\includegraphics[width = 0.47 \textwidth]{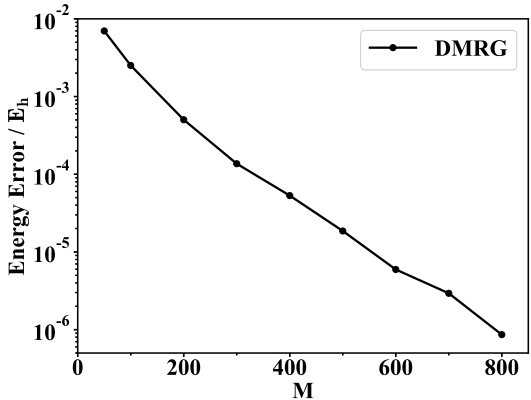}  

\caption{\label{fig:h2o}
The error of the ground state energy of H\textsubscript{2}O  with 6-31g basis calculated by MPO based DMRG algorithm with different MPS bond dimension $M_S$. The two-site algorithm is adopted to optimize the ground state MPS. The reference is the FCI energy $E_{\textrm{FCI}} = -76.11969704 \, E_\textrm{h}$. The MPO is generated by the new automatic MPO construction algorithm.
}
\end{figure}

We also calculate the ground state energy of water molecule with 6-31g basis by the MPO-based DMRG algorithm. The structure of H\textsubscript{2}O in the Cartesian coordinates is $\textrm{O}(0,0,-0.0644484)$, $\textrm{H}(\pm0.7499151,  0, 0.5114913)$ in Angstroms.
The electron integral and the reference full configuration interaction (FCI) result are calculated by PySCF.~\cite{sun2018pyscf} The DMRG results with different $M_S$ are shown in Fig.~\ref{fig:h2o}.   The error of ground state energy with $M_S = 800$ is less than $1\times 10^{-6} \, E_\textrm{h}$, which verifies the correctness of the MPO generated by the automatic algorithm.

\subsection{\add{Anharmonic vibrational Hamiltonian}}

The third example considered is the anharmonic vibrational Hamiltonian. 
There are two difficulties to solve the vibrational problems. One is how to calculate the matrix elements of high-dimensional PES as introduced in the first section. The other is how to calculate the eigenstates or simulate the dynamics. Both of these two difficulties stem from the curse of dimensionality.
To overcome the second difficulty, there have been a series of methods at different hierarchical levels including vibrational self consistent field (VSCF), vibrational perturbation theory (VPT), vibrational configuration interaction (VCI), vibrational coupled cluster (VCC) and multi-reference approaches.~\cite{barone2005anharmonic,christiansen2005beyond,christiansen2012selected,qu2019quantum,mizukami2013second,pfeiffer2014multi} (ML-)MCTDH combined with the improved relaxation algorithm~\cite{doriol2008computation} is another efficient method to obtain the eigenstates of vibrational Hamiltonian.
Recently, DMRG has also been proposed to solve the anharmonic vibrational problem.~\cite{rakhuba2016calculating,baiardi2017vibrational,baiardi2019optimization} 
Herein we use an approximate form of the Watson Hamiltonian, in which only the second-order Coriolis terms are included,~\cite{carbonniere2004coriolis,baiardi2017vibrational,sibaev2016pyvci}
\begin{gather}
    \hat{H} = \hat{H}_{\textrm{vib}} + \hat{H}_{\textrm{Cor}} \\
    \hat{H}_{\textrm{vib}} = -\frac{1}{2}\sum_{i} \frac{\partial^2}{\partial q_i^2} + V(\{\bm{q}\})    \label{eq:vH} \\
   \hat{H}_{\textrm{Cor}} = 
-\sum_{\alpha} B_{\alpha} \sum_{i<j} \sum_{k<l} \zeta_{i j}^{\alpha} \zeta_{k l}^{\alpha}\left(q_{i} \frac{\partial}{\partial q_{j}}-q_{j} \frac{\partial}{\partial q_{i}}\right) 
\left(q_{k} \frac{\partial}{\partial q_{l}}-q_{l} \frac{\partial}{\partial q_{k}}\right)
\end{gather}
$B_{\alpha}$ are the rotational constants and $\zeta_{i j}^{\alpha}$ the Coriolis coupling constants. In this numerical example, $V(\{\bm{q}\})$ is approximated as a sixth order Taylor expansion around the equilibrium geometry. 
\begin{gather}
 V(\{\bm{q}\}) = V_0 + \frac{1}{2} \sum_i \omega_i^2 q_i^2 + \frac{1}{3!} \sum_{ijk} F_{ijk} q_i q_j q_k
   + \frac{1}{4!} \sum_{ijkl} F_{ijkl} q_i q_j q_k q_l \nonumber \\
      + \frac{1}{5!} \sum_{ijklm} F_{ijklm} q_i q_j q_k q_l q_m
   + \frac{1}{6!} \sum_{ijklmn} F_{ijklmn} q_i q_j q_k q_l q_m q_n \label{eq:sexticff}
\end{gather}
It is a nontrivial task to construct a compact MPO of the operator in Eq.\eqref{eq:sexticff}, because up to six sites are coupled together in a DMRG chain, more complicated than the ab initio electronic Hamiltonian. We note that two methods have been used to construct MPO of this type of operators. In Ref.~\citenum{rakhuba2016calculating} a compact MPO is constructed by SVD compression and in Ref.~\citenum{baiardi2017vibrational} a symbolic MPO is constructed in the second quantization formalism as the electronic Hamiltonian.~\cite{keller2015efficient} 
We will use the automatic MPO construction algorithm to demonstrate its effectiveness and generality. Though we use a PES expanded as a Taylor series in this example, it is worth mentioning that the algorithm is suitable to any PES expressed as an analytical SOP form.
The molecule we choose is the widely studied C\textsubscript{2}H\textsubscript{4} molecule~\cite{baiardi2017vibrational,baiardi2019optimization, christiansen2005beyond,sibaev2015p,sibaev2016pyvci}. The PES of C\textsubscript{2}H\textsubscript{4} used here is a sextic force field as Eq.~\eqref{eq:sexticff} from PyPES library,~\cite{sibaev2015p} which is an adaptation of the PES constructed at CCSD(T) level 
with quadruple-zeta basis in internal coordinates.~\cite{delahaye2014new} 
The constant $V_0$ is set to 0 for simplicity.
Since C\textsubscript{2}H\textsubscript{4} at equilibrium geometry has D\textsubscript{2h} point group symmetry, there are only 2644 nonzero potential energy terms in the Hamiltonian otherwise it would be 18485 terms.
\begin{figure}[htbp]
\centering
\includegraphics[width = 0.47 \textwidth]{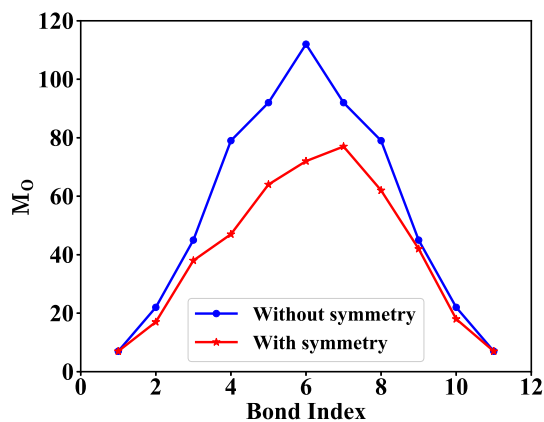}  

\caption{\label{fig:c2h4_mpo}
The MPO bond dimension $M_O$ versus bond index of C\textsubscript{2}H\textsubscript{4} described by a sextic force field with or without considering D\textsubscript{2h} point group symmetry.
}
\end{figure}
Fig.~\ref{fig:c2h4_mpo} shows the MPO bond dimension at each bond of $\hat{H}_{\textrm{vib}}$ of C\textsubscript{2}H\textsubscript{4} with or without considering the point group symmetry. The 12 vibrational DoFs within the DMRG chain are arranged according to their harmonic frequencies $\omega_i$. 
With point group symmetry, the largest MPO bond dimension is reduced from 112 to 77, which will reduce the computational cost spent in the DMRG static state or the time evolution calculations. Because the construction is automatic, the gain by utilizing symmetry to reduce the size of MPO is for free. Therefore, for Hamiltonian with negligible terms, it would be efficient to use the current algorithm to construct MPO after pre-screening the prefactors.  

\add{We use the linear response DMRG method under the Tamm-Dancoff approximation (DMRG-TDA) to calculate the vibrational excited states.~\cite{wouters2013thouless,nakatani2014linear} Compared to the other DMRG based algorithms for the high-lying excited states, such as DMRG with shift-and-invert scheme~\cite{dorando2007targeted,yu2017finding,baiardi2019optimization}, DMRG with the folded operator~\cite{baiardi2019optimization} and DMRG with projector and energy-shift~\cite{baiardi2017vibrational,larsson2019computing}, DMRG-TDA has the advantage that all the required eigenstates could be calculated in a single calculation and there is no need to track a specific state during the DMRG optimization procedure in order to avoid the root flipping problem.~\cite{hu2015excited,baiardi2019optimization} To the best of our knowledge, it is the first time that DMRG-TDA is applied to the vibrational correlation problem. In our calculation, the maximal occupation number (quanta) of each mode is limited to 6.
The ground state is calculated by state-averaged DMRG (SA-DMRG) for the lowest 9 states to make the renormalized basis more balanced for not only the ground state itself but also the excited states. Based on this ground state, we use DMRG-TDA to calculate all the eigenstates below 4000 cm\textsuperscript{-1}.  In Table~\ref{table:frequency}, we list the zero point energy, all 12 fundamental bands and the combination bands below 4000 cm\textsuperscript{-1} composed of a high-frequency C-H stretch and a low-frequency bend motion. The energy levels with or without considering the Coriolis coupling are both listed.
Each DMRG-TDA wavefunction is compressed to a rank one Hartree product state to assign the main configuration in the correlated wavefunction. The label of each normal mode follows Ref.~\citenum{delahaye2014new} and is listed in Table S1 of the supplementary material. For comparison, the VCI(8) results calculated by the PyVCI package ~\cite{sibaev2016pyvci} and the variational results reported in the literature ~\cite{delahaye2014new} are also listed.
Table~\ref{table:frequency} shows that the DMRG-TDA results have already converged within $1 \, \textrm{cm}^{-1}$ with only $M_S=50$ for the fundamental bands. For the combination bands listed, the convergence within $1 \, \textrm{cm}^{-1}$ could be reached with $M_S=100$ except the energy level $v_1 + v_8$ ($<1.5 \,\textrm{cm}^{-1}$). One of the reasons for this difference is that DMRG-TDA is a single site excitation method on a correlated ground state reference. Thus, it is more efficient to target the fundamental bands than the combination bands. More sophisticated DMRG-CISD method~\cite{wouters2013thouless}, in which the second-order tangent space of MPS is also included, could further improve the accuracy of the combination bands. 
All the energy levels below 4000 cm\textsuperscript{-1} are listed in Table S2 in the supplementary material. 
The root mean square deviation of the DMRG-TDA results ($M_S=200$) including the effect of Coriolis coupling compared to the available theoretical data in Ref.~\citenum{delahaye2014new} is $0.74 \, \textrm{cm}^{-1}$ for the fundamental bands and $9.12 \, \textrm{cm}^{-1}$ for all bands below 4000 cm\textsuperscript{-1}. We expect that if the same form of PES is used, the deviation would be smaller.
Though we choose a small molecule C\textsubscript{2}H\textsubscript{4} as a numerical example here, VDMRG is suitable for much larger molecules, such as that in Ref.~\citenum{baiardi2017vibrational} a peptide molecule is calculated.}

\begin{table*}[htbp]
\centering
\caption{The zero point energy (ZPE), twelve fundamental frequencies and eight stretch-bend combination frequencies below 4000 cm\textsuperscript{-1} of C\textsubscript{2}H\textsubscript{4} calculated by DMRG-TDA.}
\label{table:frequency}
\begin{tabular}{ccccccccccc}
    \hline
    \hline
     \multirow{3}{*}{assignment} & \multicolumn{5}{c}{without Coriolis term} & \multicolumn{5}{c}{with Coriolis term}  \\ \cmidrule(lr){2-6}\cmidrule(lr){7-11}
      & \multirow{2}{*}{harmonic} & \multicolumn{3}{c}{DMRG-TDA} & \multirow{2}{*}{VCI(8) \textsuperscript{\emph{a}} } & \multicolumn{3}{c}{DMRG-TDA} & \multirow{2}{*}{VCI(8) \textsuperscript{\emph{a}} } & \multirow{2}{*}{Ref.~\citenum{delahaye2014new}}\\
     &          & $M_S=50$ & $M_S=100$ & $M_S=200$ & &  $M_S=50$ & $M_S=100$ & $M_S=200$ & &  \\
    \hline
     ZPE         & 11164.45 & 11011.62 & 11011.62 & 11011.62 & 11011.63 & 11017.10 & 11016.95 & 11016.95 & 11016.96 & 11014.91\\
      $v_{10}$   &  824.97  &  820.25  &  820.01  &  819.99  &  820.11  &  823.69  &  823.55  & 823.53 &  823.66  &  822.42 \\
      $v_8$      &  950.19  &  926.68  &  926.35  &  926.33  &  926.45  &  935.43  &  935.21  & 935.18 &  935.31  &  934.29 \\
      $v_7$      &  966.39  &  942.03  &  941.68  &  941.65  & 941.78   &  950.89  &  950.64  & 950.61 &  950.74  &  949.51 \\
      $v_4$      & 1050.81  & 1017.81  & 1017.48  & 1017.45  & 1017.56  & 1026.05  & 1025.83  & 1025.80 & 1025.92  & 1024.94 \\
      $v_6$      & 1246.76  & 1222.41  & 1222.17  & 1222.15  & 1222.23  & 1224.91  & 1224.81  & 1224.79 & 1224.87  & 1225.41 \\
      $v_3$      & 1369.38  & 1342.26  & 1341.97  & 1341.95  & 1342.01  & 1342.94  & 1342.80  & 1342.79 & 1342.85  & 1342.46 \\
      $v_{12}$   & 1478.48  & 1438.61  & 1438.33  & 1438.31  & 1438.39  & 1441.90  & 1441.77  & 1441.76 & 1441.84  & 1441.11 \\
      $v_2$      & 1672.57  & 1623.23  & 1622.93  & 1622.90  & 1622.97  & 1625.51  & 1625.37  & 1625.34 & 1625.41  & 1624.43 \\
      $v_{11}$   & 3140.91  & 2978.69  & 2978.09  & 2978.01  & 2978.20  & 2985.73  & 2985.35  & 2985.28 & 2985.48  & 2985.38 \\
      $v_1$      & 3156.84  & 3017.93  & 3017.17  & 3017.06  & 3017.05  & 3020.17  & 3019.20  & 3019.07 & 3019.15  & 3018.99 \\
      $v_5$      & 3222.89  & 3072.75  & 3071.86  & 3071.20  & 3071.51  & 3079.63  & 3079.27  & 3079.17 & 3079.36  & 3079.86 \\
      $v_9$      & 3248.71  & 3092.41  & 3091.58  & 3091.39  & 3091.98  & 3101.40  & 3101.14  & 3101.07 & 3101.26  & 3101.69 \\
      $v_{11}+v_{10}$& & 3792.91 & 3790.82 & 3790.53 & 3791.73 & 3805.76 & 3804.14 & 3803.85 & 3805.05 & 3803.51 \\
      $v_{1}+v_{10}$ & & 3831.52 & 3828.88 & 3828.57 & 3829.80 & 3836.10 & 3834.30 & 3834.02 &  3835.17 & 3833.27 \\
      $v_5+v_{10}$   & & 3889.77 & 3886.71 & 3885.77 & 3887.13 & 3899.93 & 3897.17 & 3896.77 & 3897.92 &  \\
      $v_{11}+v_{8}$ & & 3902.58 & 3897.06 & 3897.02 & 3897.86 & 3926.18 & 3914.00 & 3914.03 & 3914.74 & 3912.73 \\
      $v_{9}+v_{10}$ & & 3910.79 & 3909.17 & 3908.85 & 3910.21 & 3922.38 & 3921.02 & 3920.79 & 3921.83 & 3921.08 \\
      $v_{11}+v_{7}$ & & 3917.10 & 3911.44 & 3911.39 & 3912.24 & 3936.35 & 3929.24 & 3929.22 & 3930.05 & 3927.84 \\
      $v_1+v_8$      & & 3936.71 & 3937.59 & 3936.20 & 3931.12 & 3952.56 & 3943.27 & 3942.03 &  3944.20 & 3946.68 \\
      $v_5+v_8$      & & 3974.11 & 3969.67 & 3970.02 & 3970.30 & 4006.88 & 4000.06 & 4000.43 & 4000.66 &  \\
    \hline
    \hline
\end{tabular} \\
\textsuperscript{\emph{a}} the VCI(8) results are calculated by Package PyVCI from Ref.~\citenum{sibaev2015p,sibaev2016pyvci}. VCI(8) means that up to 8 quanta could be excited in the CI calculation $\sum_{i}^N n_i \leq 8$ ($n_i$ is the number of quanta of the $i$th mode).
\end{table*}

\begin{figure}[htbp]
\centering
\includegraphics[width = 0.47 \textwidth]{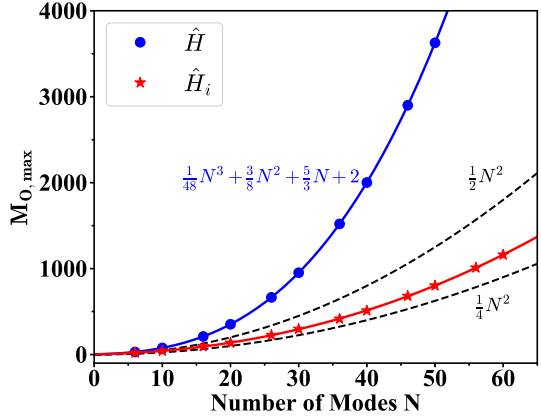}  

\caption{\label{fig:vDMRG}
The maximal MPO bond dimension $M_{O, \textrm{max}}$ of the vibrational Hamiltonian $\hat{H}_{\textrm{vib}}$ in Eq.~\eqref{eq:vH} (blue circle) and $\hat{H}_i$ in Eq.~\eqref{eq:sumofH_m} (red star) versus the number of modes.  The blue curve $\frac{1}{48}N^3 + \frac{3}{8}N^2 +\frac{5}{3}N +2$ exactly fits the $M_{O, \textrm{max}}$ of $\hat{H}_{\textrm{vib}}$. The two black dashed curves $\frac{1}{4}N^2$ and $\frac{1}{2}N^2$ indicate the scaling of $M_{O, \textrm{max}}$ of $\hat{H}_i$ is $\mathcal{O}(N^2)$ and the prefactor is between $\frac{1}{4}$ and $\frac{1}{2}$.
}
\end{figure}

Finally, we briefly discuss the computational scaling when using the MPO based DMRG algorithms. When dealing with the ab initio electronic Hamiltonian, it has been pointed out that directly treating the MPO as a dense matrix will result in an incorrect scaling $\mathcal{O}(N^5)$ compared to $\mathcal{O}(N^4)$ of the original DMRG algorithm in which only the renormalized operator matrix is retained.~\cite{keller2015efficient,chan2016matrix}
The same problem will arises for the vibrational Hamiltonian with sextic force field. 
The $M_{O,\textrm{max}}$ of $\hat{H}_{\textrm{vib}}$ with the number of vibrational modes is shown in Fig.~\ref{fig:vDMRG} (blue curve). 
For $\hat{H}_{\textrm{vib}}$, $M_{O,\textrm{max}}=\frac{1}{48}N^3 + \frac{3}{8}N^2 +\frac{5}{3}N +2$ when $N$ is even. The leading term $\mathcal{O}(N^3)$ comes from the 3-index normal operators in each block and the prefactor is $\binom{N/2}{3}$.
When calculating the expectation value $\langle \Psi |\hat{H}|\Psi\rangle$ or optimizing the ground state, the cost spent in each blocking process is $\mathcal{O}(N^6)$, because the size of each local matrix in MPO is $\mathcal{O}(N^3) \times \mathcal{O}(N^3)$. Hence, the total cost after each sweep is $\mathcal{O}(N^7)$.
However, with the original DMRG algorithm, the bottleneck in the blocking process is to contract the  3-index normal operators $q_i q_j q_k$ in the L-block (R-block) and 1-index operator $q_l$ in the center site to the complementary operator of $q_m q_n$ in the R-block (L-block), which is $\hat{P}_{mn} = \sum_{ijk} F_{ijklmn} q_i q_j q_k q_l$, with a local computational scaling $\mathcal{O}(N^5)$ and in total $\mathcal{O}(N^6)$ in one sweep. 
To recover the correct scaling in the MPO based algorithm, two approaches have been proposed. One is to fully employ the sparsity of MPO when contracting tensors.~\cite{keller2015efficient} The other method is to split the total $\hat{H}_{\textrm{vib}}$ into a sum of $\hat{H}_i$,~\cite{li2017spin,chan2016matrix} $\hat{H}_{\textrm{vib}} = \sum_{i=1}^{N} \hat{H}_i$,  where
\begin{gather}
    \hat{H}_i = -\frac{1}{2} \frac{\partial^2}{\partial q_i^2} + V_0/N + \frac{1}{2} \omega_i^2 q_i^2 + \frac{1}{3!} \sum_{jk} F_{ijk} q_i q_j q_k
   + \frac{1}{4!} \sum_{jkl} F_{ijkl} q_i q_j q_k q_l \nonumber \\
      + \frac{1}{5!} \sum_{jklm} F_{ijklm} q_i q_j q_k q_l q_m
   + \frac{1}{6!} \sum_{jklmn} F_{ijklmn} q_i q_j q_k q_l q_m q_n \label{eq:sumofH_m}.
\end{gather}

For $\hat{H}_i$ with index $i$ fixed, the maximal 5-free-index operator will give an MPO with $M_{O,\textrm{max}}= \mathcal{O}(N^2)$. The red curve in Fig.~\ref{fig:vDMRG} shows $M_{O,\textrm{max}}$ of $\hat{H}_i$ with the number of modes. The prefactor of the leading term is between $\frac{1}{4}$ and $\frac{1}{2}$. If all the $N$ sub-MPOs are added up according to Eq.~\eqref{eq:mpoadd}, the total MPO of $\hat{H}_{\textrm{vib}}$ will be recovered with the same scaling of the bond dimension $\mathcal{O}(N^3)$ but with a larger prefactor. 
The advantage to introduce $\hat{H}_i$ is that the contraction of $\hat{H}$ could be first divided into contractions of $\hat{H}_i$ and then are summed up together. Even though the MPO of $\hat{H}_i$ is treated as a dense matrix, the computational scaling in the blocking process is $\mathcal{O}(N^4)$ for each of them and the total $N$ MPOs will result in $\mathcal{O}(N^5)$. Therefore, this ``sum of MPO'' algorithm not only recovers the correct computational scaling but also is easy to be parallelized.

\section{Conclusion and Outlook}
In this work, we propose a new generic algorithm for the construction of matrix product operator of any operator with an analytical sum-of-products form based on the bipartite graph theory. 
The most important feature of the algorithm is that it could translate the operator expression to the MPO representation automatically. Therefore, it is very useful for the current (TD-)DMRG methods to be easily extended to more problems described by different Hamiltonians.
The idea of the new algorithm is to map the complementary operator selection problem to a minimum vertex cover problem in a bipartite graph, which could be elegantly solved by several well-established algorithms to get a locally optimal solution. We also prove that 
the constructed MPO is globally optimal. In addition, the new algorithm is symbolic and the sparsity of the Hamiltonian is fully preserved, which could be utilized to reduce the computational cost when contracting the tensors. We demonstrate the generality of the new algorithm by constructing MPOs ranging from the simple spin-boson model, Holstein model to the more complicated ab initio electronic Hamiltonian, and vibrational Hamiltonian described by a sextic force field. In all of the examples, the new algorithm performs well in that it could find out the small redundancy in the near-optimal hand-crafted MPO and it could take advantage of the symmetry to reduce the dimension of MPO. \add{Finally, one potential use-case of the presented algorithm not covered in the former examples is that if the coordinates system in the vibrational problem is curvilinear coordinates instead of normal mode rectilinear coordinates, the nuclear kinetic energy operators could be very complicated. The polyspherical approach developed by Gatti and co-workers,~\cite{gatti2009exact,ndong2012automatic} provides a general and analytical form of the kinetic energy operators expressed in terms of curvilinear coordinates, which often have thousands of summands and are very sensitive to numerical errors due to singularities. We expect that
the symbolic feature of the current algorithm may be useful to contract the kinetic energy operators in these cases to reduce the computational cost when calculating the matrix elements.}

\add{
\section*{supplementary material}
See the supplementary material for the label of each normal mode of C\textsubscript{2}H\textsubscript{4} and the energy levels below 4000 cm\textsuperscript{-1} calculated by DMRG-TDA.
}

\begin{acknowledgments}
J.R thank Hua Gu for insightful discussions on the graph theory and algorithm and also thank Zhendong Li for critically reading the manuscript
and helpful comments. This work was supported by the National Natural Science Foundation of China through the project “Science CEnter for Luminescence from Molecular Aggregates (SCELMA),” Grant No 21788102, as well as by the Ministry of Science and Technology of China through the National Key R\&D Plan, Grant No 2017YFA0204501. J.R. is also supported by the Shuimu Tsinghua Scholar Program.
\end{acknowledgments}

\section*{DATA AVAILABILITY}
The data that support the findings of this study are available from the corresponding author upon reasonable request.

\appendix
\section{Proof of the global optimality}
\label{sec:global_opt}
We will prove that sweeping from left to right with the procedure described in section \ref{sec:core_alg} will not change the rank of the  adjacency matrix at the same boundary. Starting from the left at the boundary between site $1$ and $2$, after the first loop through step \ref{item:step1} to step \ref{item:step3}, the coefficient matrix $\gamma_1$ is factorized as
\begin{gather}
    \gamma_1 =  \gamma_{z_1, z_2 \cdots z_N} =  \sum_{w_1} W[1]_{z_1,w_1} C[2:N]_{w_1, z_2z_3 \cdots z_N} \label{eq: matrixdecompostion1}
\end{gather}
The matrix $W[1]_{z_1,w_1}$ is the transformation matrix given in step \ref{item:step3}.
According to Lov\'asz's theorem, the dimension of $w_1$, $|w_1|$, is equal to $r_1$. Eq.~\eqref{eq: matrixdecompostion1} is nothing but a special rank decomposition of matrix $\gamma_1$. Hence, both $W[1]$ (columns are linearly independent) and $C[2:N]$ (rows are linearly independent) have rank $r_1$.  Thus,
\begin{gather}
    C[2:N]_{w_1, z_2z_3 \cdots z_N} = (W[1]^T W[1])^{-1} W[1]^T \gamma_1 = X[1] \gamma_1
\end{gather}

We will show that the unfolding matrices of $C[2:N]$, which are $C[2:N]_2=C[2:N]_{w_1 z_2, z_3 \cdots z_N}$, $C[2:N]_3=C[2:N]_{w_1 z_2z_3,  z_4\cdots z_N}$,$\cdots$, $C[2:N]_{N-1}=C[2:N]_{w_1 z_2 \cdots z_{N-1},  z_N}$ all have $\textrm{rank}(C[2:N]_i) = r_i$. 
Since $\gamma_i$ has rank $r_i$, a rank decomposition exists
\begin{gather}
    \gamma_i = \gamma_{z_1 \cdots z_i, z_{i+1} \cdots z_N}
     = \sum_{\beta=1}^{r_i} H_{z_1 \cdots z_i,\beta} F_{\beta, z_{i+1} \cdots z_N} 
\end{gather}
Thus,
\begin{align}
 C[2:N]_i & = C[2:N]_{w_1 z_2 \cdots z_i, z_{i+1} \cdots z_N}  \nonumber \\
 & = \sum_{z_1} X[1]_{w_1 z_1} \gamma_{z_1 z_2 \cdots z_i, z_{i+1} \cdots z_N} \nonumber \\
  & = \sum_{z_1} X[1]_{w_1 z_1} \big( \sum_{\beta=1}^{r_i} H_{z_1 \cdots z_i,\beta} F_{\beta, z_{i+1} \cdots z_N} \big) \nonumber \\
  & =  \sum_{\beta=1}^{r_i} \big( \sum_{z_1} X[1]_{w_1 z_1} H_{z_1 \cdots z_i,\beta} \big)  F_{\beta, z_{i+1} \cdots z_N}  \nonumber \\
  & = \sum_{\beta=1}^{r_i}  M_{w_1 z_2\cdots z_i,\beta} F_{\beta, z_{i+1} \cdots z_N} 
\end{align}
Therefore, $\textrm{rank}(C[2:N]_i) \leq r_i$. On the other hand,  if $\textrm{rank}(C[2:N]_i) < r_i$, 
\begin{align}
    \gamma_i & = \sum_{w_1} W[1]_{z_1,w_1} C[2:N]_{w_1, z_2z_3 \cdots z_N} \nonumber \\
     & = \sum_{w_1} W[1]_{z_1,w_1}  \big( \sum_{\beta=1}^{\textrm{rank}(C[2:N]_i)}   M'_{w_1 z_2\cdots z_i,\beta}  
     F'_{\beta, z_{i+1} \cdots z_N} \big) \nonumber \\
    & =   \sum_{\beta=1}^{\textrm{rank}(C[2:N]_i)} \big( \sum_{w_1} W[1]_{z_1,w_1}   M'_{w_1 z_2\cdots z_i,\beta}  \big)
     F'_{\beta, z_{i+1} \cdots z_N} \big) \nonumber \\
    & =  \sum_{\beta=1}^{\textrm{rank}(C[2:N]_i)}
    H'_{z_1 \cdots z_i,\beta}  F'_{\beta, z_{i+1} \cdots z_N}
\end{align}
The decomposition is contradictory to that the rank of $\gamma_i$ is $r_i$. Thus, $\textrm{rank}(C[2:N]_i) = r_i$.
The symbolic bipartite adjacency matrix between site 2 and 3 is $C[2:N]_2$, which has rank $r_2$. $C[2:N]_2$ could be further symbolically decomposed by finding the maximum matching and then the transformation matrix in step~\ref{item:step3}.
\begin{gather}
C[2:N]_2  =  \sum_{w_2=1}^{r_2} W[2]_{w_1z_2, w_2} C[3:N]_{w_2, z_3\cdots z_N}
\end{gather}
The process can be continued by induction. This whole proof is very similar to Theorem 2.1 of Ref.~\citenum{oseledets2011tensor}.  The difference is that the equality $\textrm{rank}(C[i:N]_j) = r_j \quad (j \geq i)$ always holds. The proof above could be intuitively understood from the fact that the rank of the coefficient matrix between two sub-systems will not change after sequential linear combinations of the basis in each sub-system as long as the new basis is linearly independent. Therefore, after sweeping from the left to the boundary between site $i$ and $i+1$, since the rank of the bipartite adjacency matrix $C[i:N]_i$ is $r_i$, the minimal number of retained operators is the same as the case that all normal operators are retained without any combination (complementary operators). As a result, the locally optimal solution is also globally optimal.

\bibliography{ref}

\end{document}


\preprint{AIP/123-QED}

\title{A General Automatic Method for Optimal Construction of Matrix Product Operators Using Bipartite Graph Theory}

\author{Jiajun Ren}
 \email{renjj@mail.tsinghua.edu.cn}
 \affiliation{MOE Key Laboratory of Organic OptoElectronics and Molecular
 Engineering, Department of Chemistry, Tsinghua University, Beijing 100084, People's Republic of China }
\author{Weitang Li}
 \affiliation{MOE Key Laboratory of Organic OptoElectronics and Molecular
 Engineering, Department of Chemistry, Tsinghua University, Beijing 100084, People's Republic of China }
 \author{Tong Jiang}
 \affiliation{MOE Key Laboratory of Organic OptoElectronics and Molecular
 Engineering, Department of Chemistry, Tsinghua University, Beijing 100084, People's Republic of China }
\author{Zhigang Shuai}%
\affiliation{MOE Key Laboratory of Organic OptoElectronics and Molecular
 Engineering, Department of Chemistry, Tsinghua University, Beijing 100084, People's Republic of China 
}%

\date{\today}

\maketitle

\subsection{Additional data on C\textsubscript{2}H\textsubscript{4}
}
\renewcommand{\arraystretch}{0.6}
\begin{table*}[bht]
\centering
\caption{Label, harmonic frequency and assignment of C\textsubscript{2}H\textsubscript{4} normal modes}
\label{table:frequency}
\begin{tabular}{ ccc }
    \hline
    \hline
label & harmonic frequency & assignment   \\
\hline
 $v_1$ &3156.84 & symmetric CH stretch \\
 $v_2$ &1672.57& CC stretch\\
 $v_3$ &1369.38& symmetric HCH bend\\
 $v_4$ &1050.81& H\textsubscript{2}C–CH\textsubscript{2} twist\\
 $v_5$ &3222.89& \textit{trans} CH stretch \\
 $v_6$ &1246.76& antisymmetric HCH wag\\
 $v_7$ &966.39& symmetric out of plane\\
 $v_8$ &950.19& antisymmetric out of plane\\
 $v_9$ &3248.71& \textit{cis} CH stretch \\
 $v_{10}$ &824.97& symmetric HCH wag\\
 $v_{11}$ &3140.91& antisymmetric CH stretch\\
 $v_{12}$ &1478.48& antisymmetric HCH bend\\
    \hline
    \hline
\end{tabular} \\
\end{table*}


\begin{ThreePartTable}
  \begin{TableNotes}
     \item[a] the VCI(8) results are calculated by Package PyVCI from Ref.~\citenum{sibaev2015p,sibaev2016pyvci}. VCI(8) means that up to 8 quanta could be excited in the CI calculation $\sum_{i}^N n_i \leq 8$ ($n_i$ is the number of quanta of the $i$th mode).
    \item[b] root mean square (rms) derivations of DMRG-TDA and VCI(8) compared to the available data in Ref.~\citenum{delahaye2014new} by variational calculation.
  \end{TableNotes}
\begin{longtable*}{cccccc}
\caption{The vibrational energy levels of C\textsubscript{2}H\textsubscript{4} below 4000 cm\textsuperscript{-1} calculated by DMRG-TDA  with $M_S=200$.} \\
    \hline
    \hline
     \multirow{2}{*}{assignment} & \multicolumn{2}{c}{without Coriolis term} & \multicolumn{3}{c}{with Coriolis term}  \\ \cmidrule(lr){2-3}\cmidrule(lr){4-6}
       & DMRG-TDA & VCI(8) \textsuperscript{\emph{a}} & DMRG-TDA & VCI(8) \textsuperscript{\emph{a}} & Ref.~\citenum{delahaye2014new}\\
    \hline
    \endfirsthead
        \hline
     \multirow{2}{*}{assignment} & \multicolumn{2}{c}{without Coriolis term} & \multicolumn{3}{c}{with Coriolis term}  \\ \cmidrule(lr){2-3}\cmidrule(lr){4-6}
       & DMRG-TDA & VCI(8) \textsuperscript{\emph{a}} & DMRG-TDA & VCI(8) \textsuperscript{\emph{a}} & Ref.~\citenum{delahaye2014new}\\
    \hline
    \endhead
    
    \hline
    \endfoot
    \hline
    \hline
    \insertTableNotes
    \endlastfoot
	 ZPE                          &   11011.62 & 11011.63  &    11016.95   &  11016.96  & 11014.91\\
   $v_{10}$               	      &     819.99 &   820.11  &	  823.53   &    823.66  &   822.42\\   
   $v_{8}$               	      &     926.33 &   926.45  &	  935.18   &    935.31  &   934.29\\
   $v_{7}$               	      &     941.65 &   941.78  &	  950.61   &    950.74  &   949.51\\
   $v_{4}$               	      &    1017.45 &  1017.56  &	 1025.80   &   1025.92  &  1024.94\\
   $v_{6}$               	      &    1222.15 &  1222.23  &	 1224.79   &   1224.87  &  1224.26\\
   $v_{3}$               	      &    1341.95 &  1342.01  &	 1342.79   &   1342.85  &  1342.46\\
   $v_{12}$              	      &    1438.31 &  1438.39  &	 1441.76   &   1441.84  &  1441.11\\
   $v_{2}$               	      &    1622.90 &  1622.97  &	 1625.34   &   1625.41  &  1624.43\\
   $2v_{10}$             	      &    1655.69 &  1655.80  &	 1661.67   &   1661.78  &  1658.39\\
   $v_{10}+v_{8}$        	      &    1748.05 &  1748.69  &	 1760.29   &   1760.92  &  1757.70\\
   $v_{10}+v_{7}$        	      &    1766.19 &  1766.84  &	 1781.02   &   1781.66  &  1778.34\\
   $v_{4}+v_{10}$        	      &    1837.67 &  1838.29  &	 1851.03   &   1851.64  &  1848.61\\
   $2v_{8}$              	      &    1858.55 &  1859.00  &	 1875.80   &   1876.24  &  1873.73\\
   $v_{8}+v_{7}$         	      &    1871.42 &  1872.13  &	 1888.43   &   1889.12  &  1885.12\\
   $2v_{7}$              	      &    1887.27 &  1887.75  &	 1904.78   &   1905.25  &  1901.61\\
   $v_{4}+v_{8}$         	      &    1939.46 &  1940.14  &	 1956.36   &   1957.02  &  1953.27\\
   $v_{4}+v_{7}$         	      &    1953.04 &  1953.74  &	 1969.93   &   1970.62  &  1966.50\\
   $v_{6}+v_{10}$        	      &    2037.41 &  2037.97  &	 2043.94   &   2044.50  &  2041.34\\
   $2v_{4}$              	      &    2032.14 &  2032.71  &	 2048.65   &   2049.22  &  2046.44\\
   $v_{6}+v_{8}$         	      &    2150.20 &  2150.72  &	 2164.70   &   2165.22  &  2163.12\\
   $v_{10}+v_{3}$        	      &    2164.53 &  2165.07  &	 2169.04   &   2169.57  &  2167.19\\
   $v_{6}+v_{7}$         	      &    2165.68 &  2166.22  &	 2177.23   &   2177.77  &  2175.40\\
   $v_{6}+v_{4}$         	      &    2239.76 &  2240.27  &	 2250.65   &   2251.16  &  2249.08\\
   $v_{12}+v_{10}$       	      &    2256.14 &  2256.69  &	 2263.44   &   2263.98  &  2260.75\\
   $v_{8}+v_{3}$         	      &    2265.28 &  2265.79  &	 2275.22   &   2275.73  &  2273.84\\
   $v_{7}+v_{3}$         	      &    2281.60 &  2282.13  &	 2291.51   &   2292.03  &  2289.89\\
   $v_{4}+v_{3}$         	      &    2355.58 &  2356.09  &	 2364.78   &   2365.28  &  2372.52\\
   $v_{12}+v_{8}$        	      &    2362.05 &  2362.60  &	 2374.53   &   2375.07  &  2363.49\\
   $v_{12}+v_{7}$        	      &    2375.37 &  2375.95  &	 2389.42   &   2389.99  &  2387.10\\
   $v_{10}+v_{2}$        	      &    2433.66 &  2434.15  &	 2439.52   &   2440.01  &  2436.64\\
   $2v_{6}$              	      &    2443.23 &  2443.66  &	 2448.56   &   2448.98  &  2447.25\\
   $v_{12}+v_{4}$        	      &    2454.54 &  2455.06  &	 2469.20   &   2469.70  &  2467.42\\
   $3v_{10}$             	      &    2498.64 &  2498.87  &	 2507.99   &   2508.19  &  2500.38\\
   $v_{8}+v_{2}$         	      &    2543.43 &  2543.94  &	 2555.74   &   2556.16  &  2553.41\\
   $v_{6}+v_{3}$         	      &    2561.75 &  2562.17  &	 2565.45   &   2565.87  &  2564.39\\
   $v_{7}+v_{2}$         	      &    2560.27 &  2560.79  &	 2571.91   &   2572.32  &  2569.48\\
   $2v_{10}+v_{8}$       	      &    2584.12 &  2585.56  &	 2599.27   &   2600.70  &  2593.75\\
   $2v_{10}+v_{7}$       	      &    2604.69 &  2606.15  &	 2624.05   &   2625.51  &  2618.33\\
   $v_{4}+v_{2}$         	      &    2637.26 &  2637.79  &	 2648.28   &   2648.69  &  2646.09\\
   $v_{6}+v_{12}$        	      &    2654.68 &  2655.11  &	 2660.96   &   2661.39  &  2659.40\\
   $2v_{3}$              	      &    2682.16 &  2682.51  &	 2683.81   &   2684.16  &  2683.05\\
   $v_{4}+2v_{10}$       	      &    2673.08 &  2674.50  &	 2690.16   &   2691.59  &  2685.01\\
   $2v_{8}+v_{10}$       	      &    2682.45 &  2684.25  &	 2702.97   &   2704.75  &  2698.44\\
   $v_{10}+v_{8}+v_{7}$  	      &    2698.17 &  2700.31  &	 2720.49   &   2722.62  &  2714.55\\
   $2v_{7}+v_{10}$       	      &    2716.76 &  2718.60  &	 2742.50   &   2744.31  &  2737.25\\
   $v_{12}+v_{3}$        	      &    2774.80 &  2775.22  &	 2779.20   &   2779.61  &  2778.00\\
   $v_{10}+v_{4}+v_{8}$  	      &    2762.06 &  2764.19  &	 2783.71   &   2785.83  &  2778.04\\
   $v_{10}+v_{4}+v_{7}$  	      &    2778.46 &  2780.60  &	 2802.44   &   2804.57  &  2826.52\\
   $3v_{8}$              	      &    2796.62 &  2797.74  &	 2821.91   &   2822.99  &  2796.68\\
   $2v_{8}+v_{7}$        	      &    2807.68 &  2809.68  &	 2832.50   &   2834.52  &  2818.31\\
   $v_{6}+v_{2}$         	      &    2830.68 &  2831.21  &	 2835.74   &   2836.24  &  2840.09\\
   $2v_{7}+v_{8}$        	      &    2823.12 &  2825.20  &	 2847.81   &   2849.91  &  2833.87\\
   $3v_{7}$              	      &    2838.14 &  2839.33  &	 2863.79   &   2864.93  &  2856.80\\
   $2v_{4}+v_{10}$       	      &    2852.93 &  2854.95  &	 2875.97   &   2877.95  &  2874.66\\
   $2v_{12}$             	      &    2869.48 &  2869.92  &	 2876.32   &   2876.75  &  2871.68\\
   $v_{6}+2v_{10}$       	      &    2867.51 &  2868.94  &	 2877.10   &   2878.55  &  2871.07\\
   $2v_{8}+v_{4}$        	      &    2868.56 &  2870.54  &	 2893.54   &   2895.52  &  2887.43\\
   $v_{4}+v_{8}+v_{7}$   	      &    2882.33 &  2884.81  &	 2907.18   &   2909.69  &  2898.17\\
   $2v_{7}+v_{4}$        	      &    2895.02 &  2897.14  &	 2919.95   &   2922.06  &  2911.66\\
   $v_{3}+v_{2}$         	      &    2957.57 &  2958.02  &	 2960.46   &   2960.90  &  2959.07\\
   $2v_{4}+v_{8}$        	      &    2951.26 &  2953.45  &	 2975.98   &   2978.16  &  2969.57\\
   $v_{11}$              	      &    2978.01 &  2978.20  &	 2985.28   &   2985.48  &  2985.38\\
   $v_{10}+v_{6}+v_{8}$  	      &    2967.91 &  2969.82  &	 2986.13   &   2988.04  &  2980.94\\
   $2v_{4}+v_{7}$        	      &    2963.31 &  2965.53  &	 2988.13   &   2990.32  &  2982.16\\
   $v_{10}+v_{6}+v_{7}$  	      &    2986.10 &  2988.03  &	 3004.22   &   3006.12  &  2999.83\\
   $2v_{10}+v_{3}$       	      &    3000.92 &  3001.77  &	 3007.75   &   3008.52  &  3002.86\\
   $v_{1}$               	      &    3017.06 &  3017.05  &	 3019.07   &   3019.15  &  3018.99\\
   $3v_{4}$              	      &    3044.49 &  3046.24  &	 3068.96   &   3070.68  &  3067.58\\
   $v_{6}+v_{4}+v_{10}$  	      &    3055.51 &  3057.45  &	 3071.70   &   3073.61  &  3064.61\\
   $v_{12}+v_{2}$        	      &    3070.40 &  3071.24  &	 3077.94   &   3078.67  &  3074.92\\
   $v_{5}$               	      &    3071.20 &  3071.51  &	 3079.17   &   3079.36  &  3079.86\\
   $v_{9}$               	      &    3091.39 &  3091.98  &	 3101.07   &   3101.26  &  3101.69\\
   $v_{10}+v_{8}+v_{3}$  	      &    3090.08 &  3092.01  &	 3103.51   &   3105.43  &  3099.87\\
   $v_{12}+2v_{10}$      	      &    3096.08 &  3097.18  &	 3104.75   &   3105.93  &  3100.04\\
   $2v_{8}+v_{6}$        	      &    3085.20 &  3086.65  &	 3109.84   &   3111.38  &  3101.69\\
   $v_{6}+v_{8}+v_{7}$   	      &    3099.48 &  3100.96  &	 3119.72   &   3121.61  &  3115.43\\
   $v_{10}+v_{7}+v_{3}$  	      &    3109.18 &  3111.11  &	 3125.17   &   3127.09  &  3121.54\\
   $2v_{7}+v_{6}$        	      &    3113.43 &  3115.09  &	 3133.45   &   3135.08  &  3129.32\\
   $v_{6}+v_{4}+v_{8}$   	      &    3163.65 &  3165.58  &	 3185.95   &   3187.85  &  3181.95\\
   $v_{4}+v_{10}+v_{3}$  	      &    3178.81 &  3180.77  &	 3193.15   &   3195.09  &  3193.74\\
   $v_{6}+v_{4}+v_{7}$   	      &    3177.54 &  3179.50  &	 3196.93   &   3198.86  &  3210.44\\
   $v_{10}+v_{12}+v_{8}$ 	      &    3182.60 &  3184.17  &	 3198.91   &   3200.45  &  3189.86\\
   $2v_{8}+v_{3}$        	      &    3194.82 &  3196.38  &	 3213.33   &   3214.89  &  3192.49\\
   $v_{10}+v_{12}+v_{7}$ 	      &    3198.97 &  3200.61  &	 3219.41   &   3221.01  &  3222.96\\
   $v_{8}+v_{7}+v_{3}$   	      &    3208.90 &  3210.91  &	 3227.07   &   3229.07  &  3214.14\\
   $2v_{2}$              	      &    3237.60 &  3237.39  &	 3243.10   &   3242.90  &  3240.19\\
   $2v_{7}+v_{3}$        	      &    3225.78 &  3227.40  &	 3244.29   &   3245.90  &  3238.79\\
   $v_{10}+2v_{6}$       	      &    3254.52 &  3256.39  &	 3264.14   &   3265.99  &  3259.58\\
   $v_{6}+2v_{4}$        	      &    3254.70 &  3256.50  &	 3273.66   &   3275.43  &  3272.06\\
   $2v_{10}+v_{2}$       	      &    3272.92 &  3272.05  &	 3279.59   &   3278.87  &  3270.53\\
   $v_{4}+v_{8}+v_{3}$   	      &    3274.73 &  3276.72  &	 3292.67   &   3294.64  &  3288.87\\
   $v_{12}+v_{4}+v_{10}$ 	      &    3273.51 &  3275.07  &	 3293.67   &   3295.20  &  3303.16\\
   $v_{4}+v_{7}+v_{3}$   	      &    3289.47 &  3291.47  &	 3307.29   &   3309.28  &  3320.87\\
   $2v_{8}+v_{12}$       	      &    3294.60 &  3293.95  &	 3315.93   &   3314.95  &  3289.05\\
   $v_{12}+v_{8}+v_{7}$  	      &    3304.29 &  3306.38  &	 3326.39   &   3328.45  &  3309.51\\
   $2v_{7}+v_{12}$       	      &    3318.78 &  3319.41  &	 3343.25   &   3343.43  &  3336.46\\
   $v_{10}+v_{8}+v_{2}$  	      &    3356.63 &  3358.63  &	 3372.34   &   3374.23  &  3367.05\\
   $2v_{4}+v_{3}$        	      &    3366.63 &  3368.39  &	 3383.97   &   3385.71  &  3384.25\\
   $v_{6}+v_{10}+v_{3}$  	      &    3380.53 &  3382.50  &	 3388.21   &   3390.17  &  3381.07\\
   $v_{8}+2v_{6}$        	      &    3372.88 &  3374.57  &	 3393.20   &   3394.88  &  3393.14\\
   $v_{10}+v_{7}+v_{2}$  	      &    3376.16 &  3378.15  &	 3394.04   &   3395.84  &  3388.90\\
   $v_{12}+v_{4}+v_{8}$  	      &    3374.66 &  3376.46  &	 3397.95   &   3399.73  &  3390.74\\
   $v_{7}+2v_{6}$        	      &    3388.68 &  3390.41  &	 3402.89   &   3404.60  &  3400.15\\
   $4v_{10}$             	      &    3391.05 &  3360.75  &	 3403.97   &   3373.38  &  3351.50\\
   $v_{12}+v_{4}+v_{7}$  	      &    3386.26 &  3388.14  &	 3410.76   &   3412.61  &  3405.46\\
   $3v_{10}+v_{8}$       	      &    3429.85 &  3437.85  &	 3448.11   &   3456.35  &  3437.52\\
   $v_{4}+v_{10}+v_{2}$  	      &    3448.87 &  3451.01  &	 3464.85   &   3466.82  &  3466.76\\
   $v_{4}+2v_{6}$        	      &    3461.04 &  3462.76  &	 3474.51   &   3476.20  &  3488.05\\
   $3v_{10}+v_{7}$       	      &    3453.30 &  3461.55  &	 3477.60   &   3486.23  &  3459.91\\
   $v_{6}+v_{12}+v_{10}$ 	      &    3469.75 &  3471.47  &	 3480.22   &   3481.97  &  3474.54\\
   $2v_{8}+v_{2}$        	      &    3471.76 &  3472.24  &	 3493.56   &   3493.94  &  3472.04\\
   $v_{12}+2v_{4}$       	      &    3469.06 &  3470.01  &	 3494.65   &   3495.51  &  3500.06\\
   $v_{6}+v_{8}+v_{3}$   	      &    3486.94 &  3488.63  &	 3502.77   &   3504.45  &  3500.53\\
   $v_{8}+v_{7}+v_{2}$   	      &    3484.87 &  3487.32  &	 3505.38   &   3507.73  &  3490.41\\
   $2v_{3}+v_{10}$       	      &    3507.45 &  3509.41  &	 3512.87   &   3514.81  &         \\
   $v_{6}+v_{7}+v_{3}$   	      &    3503.46 &  3505.17  &	 3516.16   &   3517.86  &         \\
   $2v_{7}+v_{2}$        	      &    3502.69 &  3503.76  &	 3522.77   &   3523.68  &  3516.66\\
   $3v_{10}+v_{4}$       	      &    3517.02 &  3524.56  &	 3538.72   &   3546.58  &         \\
   $2v_{8}+2v_{10}$      	      &    3520.49 &  3528.14  &	 3544.03   &   3552.00  &         \\
   $v_{8}+2v_{10}+v_{7}$ 	      &    3538.86 &  3548.24  &	 3565.73   &   3575.51  &         \\
   $v_{4}+v_{8}+v_{2}$   	      &    3553.70 &  3556.09  &	 3574.11   &   3576.38  &         \\
   $v_{6}+v_{4}+v_{3}$   	      &    3575.76 &  3577.48  &	 3587.69   &   3589.41  &         \\
   $v_{4}+v_{7}+v_{2}$   	      &    3568.71 &  3571.08  &	 3588.21   &   3590.38  &         \\
   $2v_{7}+2v_{10}$      	      &    3559.91 &  3568.40  &	 3592.58   &   3601.24  &         \\
   $v_{6}+v_{12}+v_{8}$  	      &    3580.72 &  3582.04  &	 3599.22   &   3600.50  &         \\
   $v_{12}+v_{10}+v_{3}$ 	      &    3596.23 &  3598.02  &	 3604.63   &   3606.41  &         \\
   $v_{6}+v_{12}+v_{7}$  	      &    3594.16 &  3595.50  &	 3611.21   &   3612.51  &         \\
   $2v_{3}+v_{8}$        	      &    3602.42 &  3604.22  &	 3613.39   &   3615.18  &         \\
   $v_{4}+2v_{10}+v_{8}$ 	      &    3598.98 &  3607.52  &	 3624.73   &   3633.60  &         \\
   $2v_{3}+v_{7}$        	      &    3619.90 &  3621.75  &	 3630.72   &   3632.55  &         \\
   $v_{4}+2v_{10}+v_{7}$ 	      &    3617.90 &  3626.67  &	 3647.64   &   3656.84  &         \\
   $v_{6}+v_{10}+v_{2}$  	      &    3638.75 &  3641.00  &	 3647.75   &   3649.93  &         \\
   $v_{10}+3v_{8}$       	      &    3623.14 &  3630.61  &	 3651.63   &   3659.46  &         \\
   $v_{10}+2v_{8}+v_{7}$ 	      &    3637.17 &  3647.06  &	 3666.93   &   3677.32  &         \\
   $2v_{4}+v_{2}$        	      &    3649.41 &  3651.30  &	 3668.65   &   3670.22  &  3664.37\\
   $3v_{6}$              	      &    3663.16 &  3664.66  &	 3671.25   &   3672.74  &  	      \\
   $v_{10}+2v_{7}+v_{8}$ 	      &    3655.68 &  3666.64  &	 3687.61   &   3698.89  &  	      \\
   $v_{6}+v_{4}+v_{12}$  	      &    3671.68 &  3672.98  &	 3689.06   &   3690.35  &  	      \\
   $2v_{12}+v_{10}$      	      &    3686.23 &  3688.09  &	 3697.21   &   3699.05  &  	      \\
   $2v_{3}+v_{4}$        	      &    3691.93 &  3693.74  &	 3701.94   &   3703.73  &  	      \\
   $3v_{7}+v_{10}$       	      &    3673.15 &  3682.11  &	 3709.37   &   3718.46  &  	      \\
   $v_{12}+v_{8}+v_{3}$  	      &    3696.22 &  3697.73  &	 3709.96   &   3711.40  &  	      \\
   $2v_{4}+2v_{10}$      	      &    3688.69 &  3695.96  &	 3716.87   &   3724.52  &  	      \\
   $3v_{10}+v_{6}$       	      &    3708.62 &  3716.08  &	 3721.74   &   3729.39  &  	      \\
   $2v_{8}+v_{4}+v_{10}$ 	      &    3694.18 &  3703.36  &	 3723.74   &   3733.31  &  	      \\
   $v_{12}+v_{7}+v_{3}$  	      &    3710.46 &  3712.04  &	 3725.57   &   3727.11  &  	      \\
 $v_{4}+v_{10}+v_{8}+v_{7}$       &    3711.24 &  3722.04  &	 3742.64   &   3753.81  &  	      \\
   $2v_{7}+v_{4}+v_{10}$ 	      &    3726.41 &  3736.56  &	 3760.50   &   3770.93  &  	      \\
   $v_{6}+v_{8}+v_{2}$   	      &    3752.58 &  3754.05  &	 3771.06   &   3772.41  &  	      \\
   $v_{10}+v_{3}+v_{2}$  	      &    3771.62 &  3773.90  &	 3778.23   &   3780.47  &  	      \\
   $v_{7}+3v_{8}$        	      &    3751.57 &  3759.94  &	 3784.02   &   3792.89  &  	      \\
   $v_{6}+v_{7}+v_{2}$   	      &    3770.03 &  3771.53  &	 3784.33   &   3785.63  &  	      \\
   $2v_{6}+v_{3}$        	      &    3780.61 &  3782.21  &	 3787.20   &   3788.80  &  	      \\
   $4v_{8}$              	      &    3758.89 &  3746.74  &	 3792.07   &   3780.17  &  	      \\
   $2v_{8}+2v_{7}$       	      &    3771.17 &  3777.52  &	 3802.86   &   3810.16  &  	      \\
   $v_{12}+v_{4}+v_{3}$  	      &    3787.64 &  3789.00  &	 3803.39   &   3804.72  &  	      \\
   $v_{10}+v_{11}$       	      &    3790.53 &  3791.73  &	 3803.85   &   3805.05  &  3803.51\\
   $v_{10}+2v_{4}+v_{8}$ 	      &    3775.33 &  3784.62  &	 3806.21   &   3815.78  &         \\
   $2v_{12}+v_{8}$       	      &    3790.96 &  3792.82  &	 3807.09   &   3808.91  &  3801.87\\
   $3v_{7}+v_{8}$        	      &    3784.86 &  3796.55  &	 3816.71   &   3828.71  &         \\
   $v_{6}+2v_{10}+v_{8}$ 	      &    3799.74 &  3807.50  &	 3821.23   &   3829.30  &  3816.58\\
   $2v_{12}+v_{7}$       	      &    3802.59 &  3804.53  &	 3821.64   &   3823.54  &         \\
   $v_{10}+2v_{4}+v_{7}$ 	      &    3790.18 &  3799.59  &	 3823.59   &   3833.15  &         \\
   $v_{1}+v_{10}$        	      &    3828.57 &  3829.80  &	 3834.02   &   3835.17  &  3833.27\\
   $v_{4}+3v_{8}$        	      &    3805.58 &  3814.06  &	 3838.29   &   3847.17  &         \\
   $v_{6}+2v_{10}+v_{7}$ 	      &    3820.53 &  3828.49  &	 3844.06   &   3852.31  &         \\
   $2v_{8}+v_{4}+v_{7}$  	      &    3819.13 &  3830.07  &	 3851.81   &   3863.26  &         \\
   $4v_{7}$              	      &    3822.20 &  3805.65  &	 3855.71   &   3838.99  &         \\
   $v_{6}+v_{4}+v_{2}$   	      &    3845.47 &  3847.15  &	 3858.96   &   3860.46  &         \\
   $3v_{10}+v_{3}$       	      &    3852.97 &  3857.81  &	 3863.06   &   3868.21  &         \\
   $2v_{7}+v_{4}+v_{8}$  	      &    3837.26 &  3850.43  &	 3869.38   &   3882.98  &         \\
   $v_{12}+2v_{6}$       	      &    3870.07 &  3871.39  &	 3879.29   &   3880.61  &         \\
   $3v_{7}+v_{4}$        	      &    3847.24 &  3858.40  &	 3879.62   &   3891.09  &         \\
   $v_{8}+v_{3}+v_{2}$   	      &    3874.65 &  3876.58  &	 3887.97   &   3889.77  &         \\
   $v_{12}+v_{10}+v_{2}$ 	      &    3882.39 &  3885.12  &	 3893.12   &   3895.71  &  3888.34\\
   $v_{10}+v_{5}$        	      &    3885.77 &  3887.13  &	 3896.77   &   3897.92  &         \\
   $3v_{4}+v_{10}$       	      &    3866.31 &  3873.62  &	 3898.85   &   3906.36  &         \\
   $2v_{3}+v_{6}$        	      &    3899.74 &  3901.37  &	 3904.49   &   3906.11  &         \\
   $v_{7}+v_{3}+v_{2}$   	      &    3892.36 &  3894.41  &	 3904.57   &   3906.37  &         \\
   $2v_{12}+v_{4}$       	      &    3884.61 &  3886.30  &	 3905.33   &   3906.99  &         \\
   $v_{6}+v_{4}+2v_{10}$ 	      &    3886.07 &  3893.20  &	 3906.71   &   3914.33  &         \\
   $2v_{8}+2v_{4}$       	      &    3878.88 &  3888.91  &	 3911.37   &   3921.78  &         \\
   $v_{8}+v_{11}$        	      &    3897.02 &  3897.86  &	 3914.03   &   3914.75  &  3912.73\\
   $v_{9}+v_{10}$        	      &    3908.85 &  3910.21  &	 3920.79   &   3921.83  &  3921.08\\
   $v_{8}+2v_{4}+v_{7}$  	      &    3895.18 &  3907.25  &	 3927.54   &   3939.85  &         \\
   $v_{7}+v_{11}$        	      &    3911.39 &  3912.24  &	 3929.22   &   3930.05  &  3927.84\\
   $2v_{8}+v_{10}+v_{6}$ 	      &    3906.96 &  3912.32  &	 3936.51   &   3941.91  &         \\
   $2v_{7}+2v_{4}$       	      &    3905.12 &  3916.62  &	 3937.19   &   3949.03  &         \\
   $v_{1}+v_{8}$         	      &    3936.20 &  3931.12  &	 3942.03   &   3944.20  &  3946.68\\
   $v_{10}+v_{6}+v_{8}+v_{7}$     &    3921.62 &  3929.78  &     3949.17   &  3957.97   &  3938.84\\
   $2v_{10}+v_{8}+v_{3}$ 	      &    3927.27 &  3939.11  &	 3949.46   &   3954.29  &  3936.52\\
   $3v_{10}+v_{12}$      	      &    3938.63 &  3945.15  &	 3951.89   &   3958.59  &  3941.16\\
   $v_{1}+v_{7}$         	      &    3947.25 &  3949.71  &	 3960.36   &   3961.25  &  		  \\
   $2v_{7}+v_{10}+v_{6}$ 	      &    3939.95 &  3947.15  &	 3969.43   &   3976.63  &  		  \\
   $2v_{10}+v_{7}+v_{3}$ 	      &    3955.41 &  3959.92  &	 3973.42   &   3979.89  &  		  \\
   $v_{4}+v_{3}+v_{2}$   	      &    3968.11 &  3970.05  &	 3979.55   &   3981.32  &  		  \\
   $3v_{4}+v_{8}$        	      &    3962.99 &  3972.18  &	 3995.29   &   4004.80  &  		  \\   
   $v_{6}+v_{12}+v_{3}$  	      &    3989.48 &  3990.92  &	 3996.92   &   3998.36  &  		  \\
   $v_{8}+v_{5}$         	      &    3970.02 &  3970.30  &	 4000.44   &   4000.66  &  		  \\
   fundamental bands rms\textsuperscript{\emph{b}}                            &            &           &    0.74         &   0.79         &         \\
   all bands rms\textsuperscript{\emph{b}}                            &            &           &    9.12         &          9.09  &         \\
\end{longtable*}
\end{ThreePartTable}

\bibliography{ref}